\documentclass[fleqn,usenatbib]{mnras}

\usepackage{newtxtext,newtxmath,bm}

\usepackage[T1]{fontenc}

\DeclareRobustCommand{\VAN}[3]{#2}
\let\VANthebibliography\thebibliography
\def\thebibliography{\DeclareRobustCommand{\VAN}[3]{##3}\VANthebibliography}

\usepackage{graphicx}	
\usepackage{amsmath}	
\usepackage{orcidlink}
\usepackage{times}
\usepackage{xspace}
\usepackage{comment}
\usepackage{listings}
\usepackage{nth}
\usepackage{xcolor}

\definecolor{forestgreen}{HTML}{228B22}

\definecolor{warmorange}{HTML}{E49404}

\title[Resolving kinematic dynamos in SPH MHD]{Kinematic dynamos and resolution limits for Smoothed
  Particle Magnetohydrodynamics}

\author[N. Shchutskyi et al.]{
Nikyta Shchutskyi,$^{1,2}$\thanks{E-mail: shchutskyi@lorentz.leidenuniv.nl}
Matthieu Schaller,$^{1,2}$
Orestis A. Karapiperis,$^{1,2}$
Federico A. Stasyszyn$^{3,4}$ \& \newauthor
~Axel Brandenburg$^{5,6}$
\\
$^{1}$Lorentz Institute for Theoretical Physics, Leiden University, PO Box 9506, NL-2300 RA Leiden, The Netherlands\\
$^{2}$Leiden Observatory, Leiden University, PO Box 9513, NL-2300 RA Leiden, The Netherlands\\
$^3$Instituto de Astronom\'ica Te\'orica y Experimental (IATE), CONICET – UNC, Laprida 854, X5000BGR C\'ordoba, Argentina\\
$^4$Observatorio Astron\'omico, Universidad Nacional de C\'ordoba, Laprida 854, X5000BGR C\'ordoba, Argentina\\
$^5$Nordita, KTH Royal Institute of Technology and Stockholm University, Hannes Alfv\'ens v\"ag 12, SE-10691 Stockholm, Sweden \\
$^6$The Oskar Klein Centre, Department of Astronomy, Stockholm University, AlbaNova, SE-10691 Stockholm, Sweden
}

\date{Accepted XXX. Received YYY; in original form ZZZ}

\pubyear{\the\year{}}

\begin{document}
\label{firstpage}
\pagerange{\pageref{firstpage}--\pageref{lastpage}}
\maketitle

\begin{abstract}

Understanding the origin and evolution of magnetic fields on cosmological
scales opens up a window into the physics of the early Universe.
Numerical simulations of such fields require a careful treatment to faithfully solve the equations of magneto-hydrodynamics (MHD) without introducing numerical artefacts. In this paper, we study the growth of the magnetic fields in controlled kinematic dynamo setups using both smoothed particle hydrodynamics implementations in the \textsc{Swift} code. We assess the quality of the reconstructed solution in the Roberts flow case against the reference implementation in the \textsc{Pencil} code and find generally a good agreement. Similarly, we reproduce the known features of the more complex ABC flow. 
Using a simple induction-diffusion balance model to analyse the results, we construct an 
"overwinding" trigger metric to locally detect regions
where the magnetic diffusion cannot counteract the expected induction because of limitations in the method's ability to resolve magnetic field gradients. This metric is then used to identify the necessary resolution
and resistivity levels to counteract the overwinding problem. We finally apply this metric to
adiabatic cosmological simulations and discuss the resolution requirements needed to resolve the
growth of the primordial fields without artefacts.

\end{abstract}




\section{Introduction}
\label{sec:introduction}

Astrophysical observations, such as rotation measure and synchrotron emission,
indicate the presence of magnetic fields on the scale of galaxies and the
intergalactic medium. Similarly, the absence of secondary X-ray signals from
blazars sets a lower bound on the magnetic field strength in cosmic voids 
\citep{Neronov_2010}. Magnetic fields are thus expected to be found embedded all the way
to the large-scale structure of the Universe. For comprehensive reviews, see
\cite{annurev:/content/journals/10.1146/annurev-astro-091916-055221} and
\cite{Korochkin_2021}. Despite the ubiquitous evidence of their presence, the
origin of the magnetic fields on cosmic scales is still unknown. For instance,
magnetic fields may have originated in the early Universe, emerging during
inflation, electroweak, or QCD phase transitions; see \cite{Durrer_2013}
for a review of these mechanisms. The study of such fields would thus open a
window towards understanding fundamental physics processes in the early
Universe, possibly much before nucleosynthesis. Alternatively, these
magnetic fields could have formed in astrophysical objects through battery
mechanisms \citep[e.g.][]{PhysRev.82.863,galaxies6040124,Attia_2021,MikhailovAndreasyan+2021+127+131}.

Irrespective of their origin, magnetic fields grow through gravitational
collapse and undergo exponential amplification via dynamo processes in dense
astrophysical environments such as galaxies
\citep[e.g.][]{annurev:/content/journals/10.1146/annurev-astro-071221-052807}. This
amplification erases any memory of initial conditions and ceases once the fields
become dynamically significant, influencing gas motion, i.e. saturating close to
an energy equipartition regime
\citep[e.g.][]{Ruzmaikin1988,Rogachevskii_2021}.
Studying magnetic fields as a
window onto the early Universe thus requires us to focus on regions of space
where these dynamo mechanisms are not dominant.

As such, the intergalactic medium (IGM) could serve as a reservoir for unaltered
primordial magnetic fields. However, numerical studies suggest that voids and
the IGM may be contaminated by magnetic fields expelled from dense astrophysical
objects, for instance through AGN activity in clusters. This pollution could
even reach a significant volume-filling fraction of the IGM
\citep[e.g.][]{Ar_mburo_Garc_a_2021}, potentially influencing rotation measure (RM) and
X-ray observations of magnetic fields in these regions
\citep[e.g.][]{Ar_mburo_Garc_a_2022, Bondarenko_2022}.
The recent work of \cite{Tjemsland+2024} shows that the fraction of space
filled by a strong intergalactic magnetic fields has to be at least 67\%.
This likely excludes most astrophysical production scenarios.

In order to make theoretical predictions, the large-scale evolution of magnetic
fields can be investigated using cosmological simulations. These simulations
typically employ the magnetohydrodynamics (MHD) approximation, which neglects
relativistic effects, plasma effects, and treats the gas as a compressible conducting fluid
in equilibrium. The equation of state is assumed to be an ideal gas with an adiabatic index $\gamma = 5/3$.
%

Simulations provide insight into the large-scale
distribution of magnetic fields. Combined with sub-grid models for galaxy
formation, such simulations are a powerful tool to predict the non-linear
evolution of matter and the coupling with magnetic fields; see for instance
\cite{Vogelsberger2020} for an overview of cosmological simulation methods.




Quantitatively studying magnetic fields in void environments and in the IGM
requires simulations that cover large scales for statistical accuracy while
maintaining sufficient resolution to capture processes relevant to magnetic
field evolution. In the largest cosmological simulations with box sizes of
$L_{\rm box}\sim \rm 100\,Mpc$ to $>1\,{\rm Gpc}$, the mass resolution typically ranges
from $m_{\rm gas} \simeq 10^6$ to $ 10^9~{\rm M}_{\odot}$
\citep[e.g.][]{Dolag_2009,Nelson2019,Schaye_2023} 
or, equivalently, the spatial resolution is only 1--10\,kpc in
the most dense regions \citep[e.g.][]{ Vazza_2014,Kaviraj_2017}. Given that the gas mass of a Milky Way-sized galaxy is around
$10^{11} {\rm M}_{\odot}$, each such galaxy is resolved by $10^2$ to $10^5$ gas
particles. For comparison, MHD studies of the magnetic field evolution in
isolated galaxies typically employ $\gtrsim 10^5$ resolution elements
\citep[e.g.][]{Wang_2009,Pakmor_2013,Rieder_2016,Pfrommer_2022}. Some of the magnetic field
amplification processes are thus possibly under-resolved in large-scale
simulations, which can be compensated by sub-grid processes akin to
large-eddy-simulation approximations \citep{10.1093/mnras/staa3532,Liu_2022}.

In addition to resolution, cosmological MHD simulations also deal with large
magnetic Reynolds numbers, $R_{\rm m}$ 
\begin{equation}
    R_{\rm m} = \frac{v_{\rm rms}}{\eta k_0},
    \label{Rm_definition}
\end{equation}
where $v_{\rm rms}$ is the root mean square velocity over the simulation volume,
$k_0 = 2 \pi/\lambda$ is the characteristic wavenumber at length scale $\lambda$, and
$\eta$ is the plasma resistivity.
MHD simulations of galaxy clusters suggest a good match with the observed shape of the magnetic field for turbulent resistivity values of
$\eta \approx 6\cdot 10^{27} {\rm cm}^2~{\rm s}^{-1}$ \citep{Bonafede_2011}.
Assuming a typical Virial radius of
$R_{\rm vir}\simeq 2.5 {\rm Mpc}$ and a cluster velocity of $v_{\rm vir}\sim
10^3 {\rm km}~{\rm s}^{-1}$, these values lead to magnetic Reynolds numbers
$R_{\rm m}\simeq 10^3$ to $10^4$.

However, semi-analytical dynamo studies in the
cosmological context provide a wider range of estimates for magnetic Reynolds
numbers. These studies, which consider various turbulence models and gas
densities, suggest that magnetic Reynolds numbers can range from $R_{\rm m} \geq
100$ to $2000$ \citep{Schekochihin_2005} or up to $ R_{\rm m} \sim 10^{17}$
\citep{Schober_2012}. The resolution requirements to directly simulate on
cosmological scales with such Reynolds numbers is much beyond current and
future computing capabilities, and here again, sub-grid process are often
employed to model the unresolved part of the turbulence cascade. Alternatively,
some simulations consider neglecting sub-grid effects and restrict their analysis
to the well-resolved regime \citep{Marinacci_2015, Mtchedlidze_2022}.

The MHD differential equations can be solved numerically using various methods. Most
simulations employ mesh-based approaches, such as \textsc{Arepo},
\citep{Pakmor_2011} 
and \textsc{Enzo} \citep{Bryan_2014}. Alternatively, meshless schemes like meshless finite mass (MFM) in \textsc{Gizmo} \citep{Hopkins_2015}, SPH-based MHD
approaches such as the ones in the \textsc{Gadget} code \citep{Dolag_2009,
  Stasyszyn_2012}, \textsc{Gasoline}  \citep{Wissing_2020} or SPH MHD implementation in \textsc{Gizmo} \citep{Hopkins_2015} have also been used for astrophysical and
cosmological simulations.

In this study, we make use of \textsc{Swift}, an open-source SPH-based code for
large-scale cosmological simulations \citep{Schaller_2024}, offering several
models for hydrodynamics and sub-grid physics. The code also includes two
variants for SPMHD: one based one the direct integration of $\textbf{B}/\rho$
\citep{Karapiperis2025}, inspired by the implementation of \cite{Price_2018}, and
the other using a vector potential approach, $\textbf{B} = {\rm curl} \, \textbf{A}$,
derived from the work of \cite{STASYSZYN2015148}. Having several independent
numerical results with different hydrodynamics models, sub-grid models, and two
MHD implementations can be advantageous for estimating the numerical uncertainty
of the results.


In this work, both MHD flavors provided in \textsc{Swift} will be tested on a
set of standard kinematic large-scale dynamo problems: the Roberts flow I
\citep{10.1098/rsta.1972.0015}, which has been studied both analytically and
through numerical simulations \citep{Tilgner_2008, STASYSZYN2015148,
  Clarke_2020}, as well as the ABC flow \citep{Teyssier_2006,
  Bouya2012RevisitingTA, Archontis_2002, Baggaley_2009}. Convergence against the
reference implementation in the \textsc{Pencil Code} \citep{2021JOSS....6.2807P} 
will be used to assess the correctness of our SPH solver. To ensure the validity
of the results, MHD simulations are expected to maintain a solenoidal magnetic
field. To monitor this, local divergence error metrics are used and corrective
measures are employed. We study the impact of such measures on the growth of the
magnetic field in these well-controlled experiments. The lessons learned from
studying these flows allow us to construct an ``over-winding'' trigger detecting
where the simulation's resolution is too low for a given kinematic dynamo to be
properly resolved. This trigger can then be used in a cosmological setting to
assess whether the geometry of the field in void regions is unaffected by
numerical artefacts and could thus be used, in the future, to put constraints on
the origin of the fields.

This paper is structured as follows. The codes, relevant differential equations
and the error metrics are described in Sec.~\ref{sec:methods}. The performance
of our simulation code on the Roberts and ABC kinematic dynamo problems is
presented in Sec.~\ref{sec:flows}. The construction of the overwinding trigger
and tests are presented in Sec.~\ref{sec:overwinding} with applications to a
simple cosmological test case following in Sec.~\ref{sec:CosmoVolume}. Finally,
in Sec.~\ref{sec:conclusions} we offer some conclusions and outlook on future
applications of our code.

\section{Methods}
\label{sec:methods}



In this section, we introduce the equations of MHD and their numerical
implementations in the \textsc{Swift} (Sec.~\ref{ssec:swift_mhd}) and
\textsc{Pencil} (Sec.~\ref{ssec:pencil}) codes. We also introduce in
Sec.~\ref{ssec:error_metrics} the error metrics we will use to monitor the
divergence errors appearing in the code.

\subsection{MHD}
The evolution of magnetic field and plasma can be modeled using the MHD
approximation \citep{RevModPhys.74.775,Brandenburg_2005}:

\begin{equation}
    \frac{\partial\textbf{B}}{\partial t} = {\rm curl} [\textbf{v}\times \textbf{B}] + \eta \Delta \textbf{B},
\label{eq:InductionEquation}
\end{equation}
\begin{equation}
    \frac{\partial\textbf{v}}{\partial t} + (\textbf{v} \cdot \nabla) \textbf{v}  = -\frac{1}{\rho} \nabla P -\frac{1}{\rho}\nabla \boldsymbol{\mathsf{S}} +\bm{\Pi},
    \label{eq:FluidEquation}
\end{equation}
with $\textbf{v}$ the fluid velocity, $\rho$ the fluid density, $\textbf{B}$ the
magnetic flux density, $\eta$ the physical resistivity, $\boldsymbol{\mathsf{S}}$ the Maxwell
stress tensor:
\begin{equation}
    S_{ij} =
    \frac{1}{2} \frac{B^2}{\mu_0} \delta_{ij}
    - \frac{B_{i} B_{j}}{\mu_0},
\end{equation}
with $\mu_0$ the vacuum permeability, and thermal pressure:
\begin{equation}
  P = (\gamma - 1) \rho u
\end{equation}
where $u$ is the specific internal energy.
The vector $\bm{\Pi}$ corresponds to an additional artificial viscosity
and will be explained below.
Additionally, there is an equation governing the specific energy density \citep[see
  e.g.][]{Price_2012}:
\begin{equation}
   \frac{\partial e}{\partial t} + (\textbf{v}\cdot {\nabla})e = -\frac{1}{\rho} \nabla_{j} v_{i} S_{ij} + \overline{\bm{\Pi}}
\end{equation}
where $e = \textbf{v}^2/2+u+\textbf{B}^2/2 \rho$ is the specific energy density, $\overline{\bm{\Pi}}$ -- energy dissipation term from viscous and resistive heating.

These equations describe the magnetic
field, velocity and specific energy evolution as a function of spatial coordinates and time in the
Eulerian frame.

\subsection{MHD in the \textsc{Swift} code}
\label{ssec:swift_mhd}


The set of equations introduced above can be discretised in the Lagrangian frame
using a particle-based approach within the framework of SPH \citep[see
  e.g.][]{Price_2012}. More specifically, our implementation is built on top of
the SPHENIX \citep{Borrow_2021} formulation of SPH that was
designed specifically to perform well in galaxy formation simulations\footnote{For completeness, all the runs employed  version \texttt{3ea21e98} of the code.}.

\subsubsection{Momentum equation in SPH}

We start with the discretisation of the momentum equation
\ref{eq:FluidEquation}. The \textsc{Swift} code discretises the gas into a set of
particles and solves the equations of motion at particle positions, i.e. in a
frame where the observer moves with the fluid:
\begin{equation}
    \frac{d\textbf{v}}{d t} =  -\frac{1}{\rho} \nabla \boldsymbol{\mathsf{S}}+\bm{\Pi}+\textbf{M}\label{FluidEquationLagrangianFrame}
\end{equation}
with $d/dt$ the material derivative, $\bm{\Pi}$ corresponding to additional artificial viscosity
terms that help capture shocks at both 
hydrodynamic and magnetic wave
discontinuities. This is achieved by using SPHENIX viscosity terms with a signal velocity that incorporates the Alfvén speed \citep{Karapiperis2025}. The term $\textbf{M}$ is an additional force that corrects
tensile instability akin to the \cite{POWELL1999284} term, an effective
numerical force that arises due to the SPH discretisation and counteracts the particle clumping associated with $\frac{1}{\rho}\nabla S$. The
magnitude of this effective force is proportional to the divergence of the magnetic
field \citep{Børve_2001}. The corrective term $\textbf{M}$ explicitly violates
energy conservation \citep{Price_2012}. The force becomes significant in regions where the magnetic pressure is comparable in magnitude to the thermal pressure. Consequently, maintaining a small divergence of the magnetic field is crucial to minimize both the particle-clumping component of $\frac{1}{\rho} \nabla S$ and the corrective term $\textbf{M}$.

In SPH, the local matter density is computed as a sum:
\begin{equation}
    \hat{\rho}_{a} = \sum_{b} m_{b} W(|\textbf{r}_{ a}-\textbf{r}_{b}|, h_{a}),
\end{equation}
where $a$ and $b$ are particle labels, $W$ is the smoothing kernel and $h$ the
smoothing length, which is related to the local mean inter-particle separation.


The MHD momentum equation becomes
\begin{multline}
\frac{d v_{a}^i}{dt} = -\sum_{b} m_{b} \Bigg[  \frac{f_{a b}S_{a}^{ij}}{\hat{\rho}_{a}^2} \frac{\partial W_{ab}(h_{a})}{\partial x_{a}^{j}}+
     \frac{f_{ba}S_{\rm b}^{ij}}{\hat{\rho}_{b}^2} \frac{\partial W_{ab}(h_{b})}{\partial x_{b}^{j}}\Bigg] + \\
     +\Pi_{a}^{ i}+M_{a}^{i}+f_{\rm grav}^{a, i},
\label{SPH_momentum_equation}
\end{multline}
where $f^{a, i}_{\rm grav}$ are the accelerations coming from gravity and $
f_{ab} $:
\begin{equation}
    f_{ab} = 1 + \frac{h_{a}}{3 \hat{\rho}_{ a}}\frac{\partial \hat{\rho}_{a}}{\partial h_{a}}.
\end{equation}
are terms accounting for the spatial variation of the smoothing lengths.  This
set of equations is identical to the pure hydrodynamical case \citep[see
  e.g.][]{Price_2018, Schaller_2024} but with the pressure replaced by the
Maxwell tensor and the tensile correction added.


\subsubsection{Magnetic field evolution in SPH using direct induction} \label{ssec:ODI}
The simplest way to implement the evolution of the magnetic fields into SPH is
to trace their evolution at particle positions:
\begin{equation}
    \frac{d\textbf{B}}{dt} = (\textbf{B}\cdot \nabla)\textbf{v} - \textbf{B}\cdot{\rm div}\textbf{v} + \eta \Delta \textbf{B} + \Omega_{\rm AR}
\end{equation}
with $\textbf{v}$ the fluid particle velocity, $\Omega_{\rm AR}$ an
artificial resistivity corrective term to aid handling of magnetic field
discontinuities \citep[see e.g.][]{Price_2012, Price_2018}.

Our direct induction (DI) implementation is fully described by
\cite{Karapiperis2025}. In summary, the equations of resistive MHD are solved by
evolving the quantity $\textbf{B}/\rho$, with $\textbf{B}$ the magnetic flux density
and $\rho$ the mass density. The induction equation then reads:

\begin{equation}
    \frac{d}{dt} 
    \left( \frac{\textbf{B}}{\rho} \right) =
    \left( \frac{\textbf{B}}{\rho} \cdot \nabla \right) \textbf{v} 
    + \frac{\eta}{\rho} \Delta \textbf{B}
    + \Omega_{\rm AR} 
    - \frac{1}{\rho} \nabla \psi
\end{equation}

\noindent with $\psi$ a scalar field (the Dedner field), which is used to remove
via hyperbolic and parabolic divergence cleaning any non-zero ${\rm div}
\textbf{B}$ arising due to numerics \citep[see e.g.][]{DEDNER2002645, TRICCO2016326}. The evolution
equation for the scalar field itself is given by:
\begin{equation}
    \frac{d}{dt} 
    \left( \frac{\psi}{c_{\rm h}} \right) =
    - \sigma_{\rm h} c_{\rm h} {\rm div} \textbf{B}
    - \frac{1}{2} \frac{\psi}{c_{\rm h}} {\rm div} \textbf{v}
    - \sigma_{\rm p}\frac{\psi}{\tau_{\rm c}}
    \label{eq:Dedner}
\end{equation}
\noindent with $c_{\rm h}$ an appropriately chosen cleaning speed and $\tau_{\rm c} $ is
local Dedner scalar dissipation time:
\begin{equation}
    c_{\rm h} = \sqrt{v_{\rm A}^2+c_{\rm S}^2}, \hspace{0.5cm} \tau_{\rm c} = \frac{h}{c_{\rm h}}, \hspace{0.5cm} v_{\rm A} =\sqrt{B^2/(\mu_{\rm 0} \rho)}
\label{cleaning_parameters}
\end{equation}
where $\sigma_{\rm h}$ and $\sigma_{\rm p}$ are constant parameters with typical values
$\sigma \simeq 1.0$ \citep{Price_2018}, $v_{\rm A}$ is the Alfv\'en speed and
$c_{\rm S}$ the sound speed.

Several ways of expressing a divergence operator in SPH exist
\citep{Price_2012}. In our implementation of the cleaning terms, we choose the
anti-symmetric divergence formulation:
\begin{equation}
{\rm div} \textbf{B}_{a} = \frac{1}{\rho_{a}}\sum_{b} m_{b}
(\textbf{B}_{b}-\textbf{B}_{a})\cdot \nabla W_{ab}(h_{a})
\end{equation}

The discretised version of the induction equation then becomes
\citep{Karapiperis2025, Price_2018}:

\begin{equation}
    \frac{d}{dt}\left( \frac{\textbf{B}}{\rho} \right)_{a} = \Omega_{\rm Str}^a + \Omega_{\rm Dender}^a + \Omega_{\rm Ohm}^a +\Omega_{\rm AR}^a 
\end{equation}
with the stretching source term:
\begin{equation}
    \Omega_{\rm Str}^a = -\frac{f_{ab}}{\hat{\rho}_{a}^2}\sum_{b} (\textbf{v}_{a}-\textbf{v}_{b})(\textbf{B}_{a} \cdot \nabla_{a}W_{ab} (h_{a}))
\end{equation}
divergence cleaning term:
\begin{equation}
   \Omega_{\rm Dender}^a = -\sum_{b} m_{b} \left[ \frac{f_{ab}\psi_{a}}{ \hat{\rho}_{a}^2}\nabla_{ a}W_{ ab} (h_{a}) + \frac{f_{ ba}\psi_{b}}{ \hat{\rho}_{b}^2}\nabla_{b}W_{ab} (h_{b})\right] 
   \label{Dedner_cleaning_source}
\end{equation}
with the Ohmic (or physical) resistivity term:
\begin{equation}
    \Omega_{\rm Ohm}^a = 2 \eta \sum_{b} m_{b} \frac{\partial_{\rm r} W_{ab}(h_{a})}{|r_{ab}|\hat{\rho}_{a}\hat{\rho}_{b}}(\textbf{B}_{a}-\textbf{B}_{b}).
\end{equation}

Note that our implementation also uses artificial resistivity terms but these
were switched off in what follows as we are concerned with the precise effect of
physical Ohmic resistivity alone.

\subsubsection{Vector Potential (VP) MHD implementation in \textsc{Swift}}

Alternatively, the magnetic field divergence constraint can be enforced by
evolving for the vector potential  \citep[VP;][]{STASYSZYN2015148} instead of the
magnetic field. The magnetic field in terms of the vector potential $\textbf{A}$ is
then:
\begin{equation}
    \textbf{B} = {\rm curl} \textbf{A}.
    \label{MF_from_VP}
\end{equation}
\noindent The induction equation for $\textbf{A}$ reads:
\begin{equation}
    \frac{d}{dt} 
    \textbf{A} = \textbf{v} \times {\rm curl} \textbf{A} + (\textbf{v} \cdot \nabla) \textbf{A} + \eta \Delta \textbf{A} - \nabla \Gamma,
\end{equation}
\noindent where $\Gamma$ is the electromagnetic gauge.
The gauge does not influence the magnetic field and only governs the evolution of the vector potential.
Since Maxwell’s equations, when expressed in terms of the vector
potential, do not uniquely determine it, a gauge condition must be chosen
\citep[][]{jackson_classical_1999}.
One possible choice is the Coulomb gauge, where $\text{div} \textbf{A} = 0$.
To enforce this condition numerically, a new scalar field can be
introduced, sourced by the residual $\text{div} \textbf{A}$ \citep[][]{Stasyszyn_2012}.
Similarly to Dedner divergence cleaning, the electromagnetic gauge
$\Gamma$ propagates and removes the divergence of the $\textbf{A}$
field through its evolution equation:
\begin{equation}
\frac{d}{dt} \Gamma = -c_{\rm h}^2(\nabla\cdot\textbf{A})-\frac{c_{\rm h}^2}{\tau_{\rm c}}\Gamma,
\end{equation}
\noindent where $c_{\rm h}$,$\tau_{\rm c}$ defined the same way as for
Dedner cleaning (Eq. \ref{cleaning_parameters}). The SPH version of the induction
equation for vector potential then reads:
\begin{equation}
\frac{d}{dt} \textbf{A}_{a} = \Omega_{\rm Str}^a+\Omega_{\rm Gauge}^a+\Omega_{\rm Ohm}^a
\end{equation}
with stretching term: 
\begin{equation}
    \Omega_{\rm Str}^a =\frac{f_{ab}}{\hat{\rho}_{a}} \sum_{b} m_{b} (\textbf{v}_{a}-\textbf{v}_{b})\textbf{A}_{a}\nabla_{a}W_{ab} (h_{a})
\end{equation}
gauge term
\begin{equation}
    \Omega_{\rm Gauge}^a = \frac{f_{ab}}{\hat{\rho}_{a}} \sum_{b} m_{b} (\Gamma_{a}-\Gamma_{b})\nabla_{a}W_{ab} (h_{a})
    \label{Gauge_cleaning_source}
\end{equation}

\noindent with the observed B field reading:
\begin{equation}
    \textbf{B}_{a} = \sum_{b} \frac{m_{b}}{\hat{\rho}_{b}}(\textbf{A}_{a}-\textbf{A}_{ b})\times\nabla_{a}W_{ab} (h_{a})).
\end{equation}

As was the case for the direct induction method, we add an Ohmic
resistivity term which, here, acts directly on $\textbf{A}$:
\begin{equation}
    \Omega_{\rm Ohm}^a = 2 \eta \sum_{b} \frac{m_{b}\hat{\rho}_{a}}{|r_{ab}|\hat{\rho}_{ab}^2} \frac{\partial_{\rm r} W_{ab}(h_{a})+\partial_{\rm r} W_{ab}(h_{b})}{2}(\textbf{A}_{a}-\textbf{A}_{b}),
\end{equation}
where $\hat{\rho}_{ab} = \frac{1}{2}(\hat{\rho}_{
  a}+\hat{\rho}_{b})$.

In what follows, we will make use of both flavours of SPMHD with
special attention to the direct induction scheme.

\subsection{MHD in the \textsc{Pencil Code}}
\label{ssec:pencil}

As mentioned above, we make use of the \textsc{Pencil} code
\citep{BRANDENBURG2002471,Brandenburg_2003}, a thoroughly tested and validated
grid code, as a reference against which to evaluate performance of
\textsc{Swift}'s MHD implementations.

In the \textsc{Pencil Code}, space is divided into a grid of fixed points. Spatial
derivatives are calculated as combination of values of a function at the point
and its neighbours in the direction of derivative using centered finite
differences approach:
\begin{equation}
f_i' = (-f_{ i-3}+9f_{ i-2}-45f_{i-1}+45f_{ i+1}-9f_{ i+2}+f_{i+3})/(60 \delta x),
\label{Pencil_df}
\end{equation}
\begin{multline}
f_i'' = (2f_{ i-3}-27f_{ i-2}+270f_{ i-1}-490f_{ i}+\\
+270f_{ i+1}-27f_{ i+2}+2f_{ i+3})/(180 \delta x^2).
\label{Pencil_d2f}
\end{multline}
For time integration the third order $2N$-Runge-Kutta scheme is
used \citep{Williamson80}.

The equations for density and velocity fields in the Eulerian frame:
\begin{align}
    \frac{\partial}{\partial t}{\rm ln} \rho + (\textbf{v}\cdot \nabla){\rm ln} \rho &= -{\rm div} \textbf{v}, \\
   \frac{\partial}{\partial t}\textbf{v}+ (\textbf{v}\cdot \nabla)\textbf{v} &= -\frac{\nabla P}{\rho}+{\rm ln}\rho
+ \frac{1}{\rho} \textbf{J}\times\textbf{B}
+\textbf{f}_{\rm visc}+\textbf{f}_{\rm grav},
\label{Pencil_u}
\end{align}
where $\textbf{f}_{\rm visc}$ is the viscous force per unit mass, which is proportional to the
viscosity $\nu$ and the divergence of the traceless rate-of-strain tensor.
The magnetic field equations are solved using the vector potential in the Weyl
gauge ($\Gamma = 0$) in an Eulerian frame:
\begin{equation}
\frac{\partial\textbf{A}}{\partial t}=\textbf{u}\times\nabla\times\textbf{A}
+\eta\nabla^2\textbf{A}.
\label{Pencil_A}
\end{equation}
To deal with discontinuities and discretization errors the code involves some
non-zero $\nu$ and $\eta$ in Eqs.~(\ref{Pencil_u}) and (\ref{Pencil_A}) using a high-order
spatial discretisation as in Eqs.~(\ref{Pencil_df}) and (\ref{Pencil_d2f}).
For the kinematic dynamo problems such as Roberts Flow I, the momentum equation is not solved explicitly.
Instead the velocity field is given as a coordinate-dependent function.
For internal energy the isothermal equation of state was used.


\subsection{Magnetic field error metrics in \textsc{Swift}}
\label{ssec:error_metrics}

In a direction induction method, magnetic field evolution divergence errors can
occur due to numerical errors. Even though the Dedner cleaning scheme acts to
reduce the divergence, it is useful to have a set of tools to monitor the
spurious monopole component of the magnetic fields. We introduce the ones we use
here.

The most widely used error metric for SPMHD is the ratio \citep[e.g.\ Eq.~(78)
  of][]{10.1111/j.1365-2966.2005.09576.x}:
\begin{equation}
    R_{0,a} =\frac{|{\rm div} \textbf{B}_{a}|\cdot h_{a}}{|\textbf{B}_{a}|}.
    \label{error_R0}
\end{equation}
Whilst directly related to the problem we want to monitor and to the term
sourcing the Dedner scalar evolution, this error metric is large only if the
divergence is of order of largest resolvable gradient by SPH, $\sim |\textbf{B}_{a}|/h_{a}$ (see below). 

In addition, the outcomes of divergence presence are not limited to some unphysical field fraction in the magnetic field, and thus require additional monitoring techniques.
If the error metric $R_0$ is small, it does not guarantee that the effect of the spurious divergence is small. For example, there are monopole forces
proportional to ${\rm div}\textbf{B}$ that could result in major forces
affecting the dynamics of a particle. To monitor such situations, another, more
rarely used metric monitors the component of the magnetic field that is parallel
to the total magnetic force acting on the fluid \citep[see Eq.~(79)
  of][]{10.1111/j.1365-2966.2005.09576.x}. It thus is a spurious force:
\begin{equation}
  R_{1,a} = \frac{(\textbf{B}_{a}\cdot \textbf{f}_{{\rm mag},a} )}{|\textbf{B}_{a}||\textbf{f}_{{\rm mag},a}|},
  \label{error_R1}
\end{equation}

where $\textbf{f}_{{\rm mag},a}$ is the sum of the first term corresponding
to $-\frac{1}{\rho}\nabla S$ in Eq.~(\ref{SPH_momentum_equation}) and the
monopole correction force $\textbf{M}_a$.

Finally, even when magnetic forces parallel to the magnetic field are small
(i.e.\ $R_1$ is small), the monopole component of the magnetic field can create
an additional spurious Lorentz force. To monitor this, one can estimate the
magnitude of the monopole component relative to the physical magnetic field. To this
end, we introduce the ratio of the magnetic field divergence to the current:
\begin{equation}
  R_{2,a} =\frac{|{\rm div}\textbf{B}_{a}|}{|{\rm curl}\textbf{B}_{a}| }.
  \label{error_R2}
\end{equation}
 
All these error metrics can be large in regions where the values of the
${\rm div}\textbf{B}_{a}$ and ${\rm curl}\textbf{B}_{a}$ operators are very
sensitive to the particle arrangement. The particle noise level can be
conservatively estimated as the level of error of the SPH gradient operator
\citep[e.g.][]{Price_2012, Violeau_book}. Assigning a unity scalar quantity to SPH gas
particles and computing a gradient vector, which in the continuum limit should be
zero, yields:
 \begin{equation}
     \langle \nabla 1 \rangle_{a} = \sum_{b} \frac{m_{b}}{\hat{\rho}_{b}} \nabla W(|\textbf{r} - \textbf{r}_{b}|, h(\textbf{r}))\bigg|_{\textbf{r}=\textbf{r}_{a}}.
 \end{equation}
This quantity will be non-zero due to particle arrangement in space. Using this
construction, the errors induced by particle noise in the divergence and curl
operators can be estimated as:
\begin{align}
  \langle \delta  {\rm  div}\textbf{B}_{a}\rangle  &\simeq ( \langle \nabla 1 \rangle_{a} \cdot \textbf{B}_{a}), \notag \\
  \langle \delta  {\rm  curl}\textbf{B}_{a}\rangle &\simeq [ \langle \nabla 1 \rangle_{a} \times \textbf{B}_{a}], \notag \\
  \langle \delta  \textbf{f}_{{\rm mag},a} \rangle       &\simeq \frac{S_{a}^{ii}}{3\hat{\rho}_{a}} \cdot \langle \nabla 1 \rangle_{a}.
  \label{errors_noise_cut}
\end{align}

In what follows, and more generally in \textsc{Swift}, when measuring the three error
metrics introduced above, we explicitly zero the quantities that are not
$10\times$ larger than their corresponding pure-noise counterparts. This allows
us to focus our analysis on regions where the errors are not triggering solely
due to the noise in the particle distribution. Note however that the Dedner
evolution equation (\ref{eq:Dedner}) does not use this limit; any (spurious)
$\rm{div}\textbf{B}$ appearing in the fluid is sourcing the scalar field evolution.

\section{Kinematic dynamo tests}
\label{sec:flows}
\begin{figure*}
    \centering
    \includegraphics[width=\linewidth]{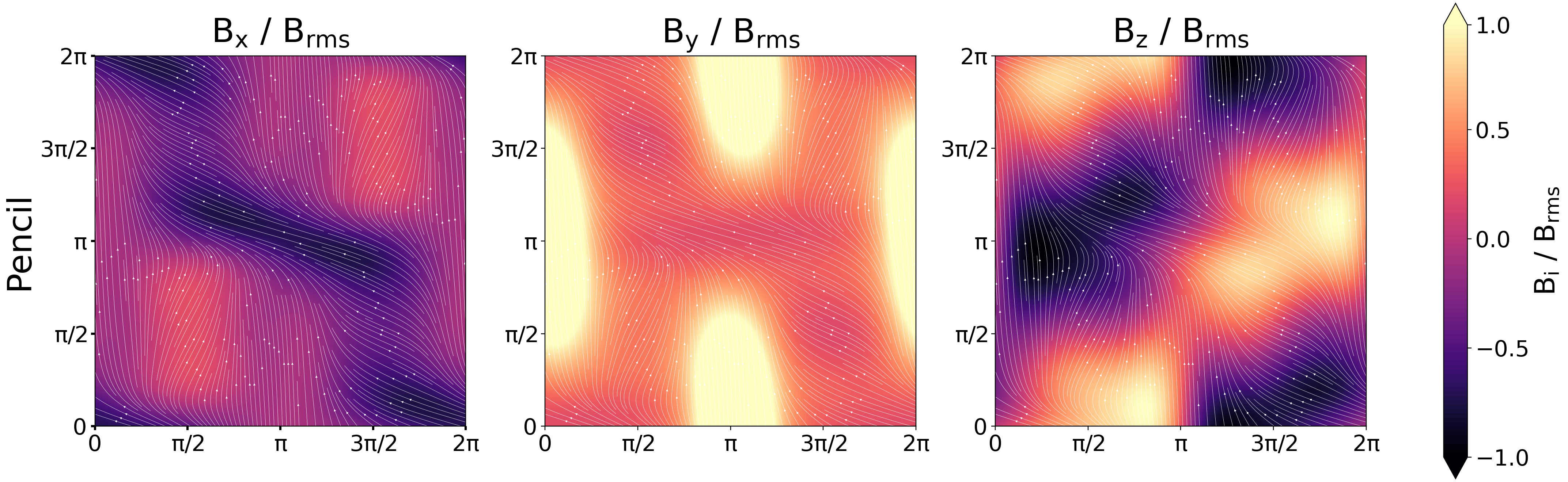}
    \includegraphics[width=\linewidth]{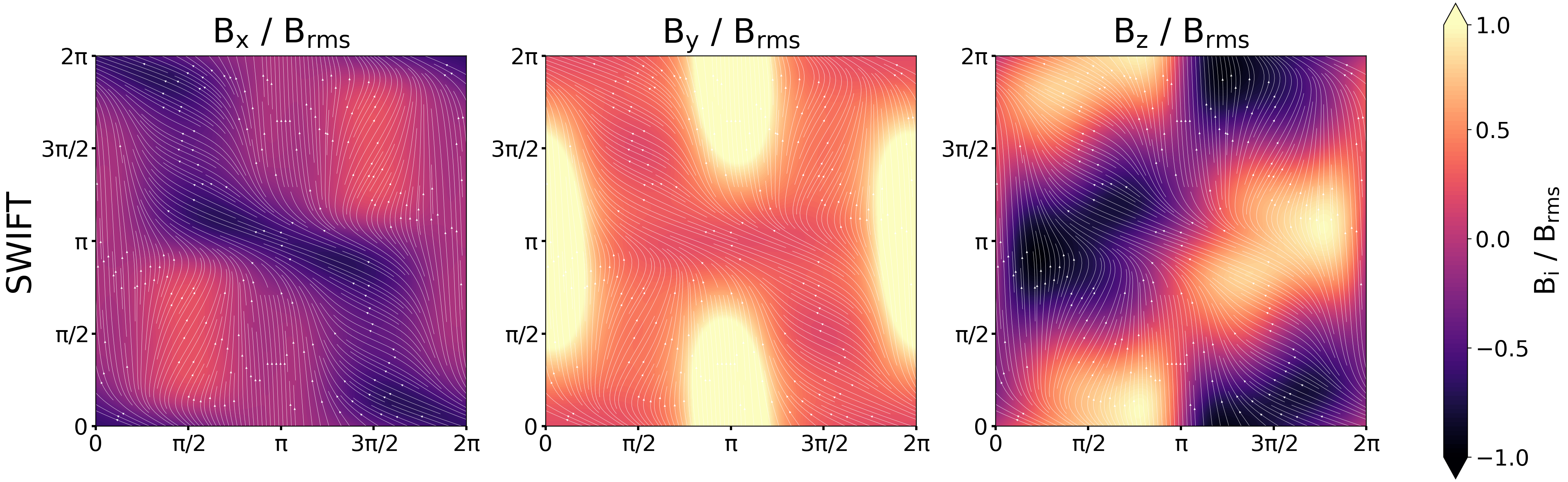}
    \caption{Maps of the magnetic field components $B_{x}$ (left), $B_{y}$ (middle), and $B_{z}$ in units of the magnetic field RMS in the
      $xy$-plane at $z=0.0$ for \textsc{Pencil} and $z=0.05$ for  \textsc{Swift} (for explanation of such choice see text) at around $t>20$ using the direct induction scheme (bottom row)
      with $N=32^3$ particles. The white
      streamlines indicate the magnetic fields inside the plane. The growing
      mode of the magnetic fields in \textsc{Swift} reproduces well the observed
      features in the fields extracted from the \textsc{Pencil} code.}
    \label{fig:RF1_field_configuration}
\end{figure*}

Dynamos are crucial for cosmological and astrophysical magnetic field evolution.
Therefore, correctly capturing the amplification process is a necessary step in
the validation of a numerical code. In this section we study the performance and
limitations of the SPMHD implementations in \textsc{Swift} code introduced above
on the standard \cite{10.1098/rsta.1972.0015} flow and the
Arnold–Beltrami–Childress \citep[ABC;][]{Arnold_2014,Childress_1970} flow
dynamo test problems. Both of these tests probe the kinematic dynamo regime.

The main aim of these test is to verify that the induction source term together
with resistive terms in the code are modelled correctly and allow correct
reproduction of magnetic field growth. In particular, we focus on
characteristics such as the growth rate dependence on the physical resistivity
and resolution convergence as well as on qualitative features such as the spacial
magnetic field distribution and the mode transitions. We compare our results to
the \textsc{Pencil} code and to published solutions.

\subsection{The Roberts flow I test}
\label{ssec:RF1}

There are several simple dynamo capable flows consisting of four vortices in the
$xy$-plane with a uniform velocity field in the $z$ direction
\citep{10.1098/rsta.1972.0015}. These flow problems have been widely studied
both with simulations and semi-analytically. The different flows proposed in
this paper vary in their $v_{\rm z}(x,y)$ dependence and provide a range of
mechanisms for magnetic field growth. One of them, the first flow, exhibits
growth via the alpha effect \citep{Tilgner_2008}. Flows II and III show growth
due to a memory effect \citep{Rheinhardt_2014}. The fourth flow provides growth
through effective negative turbulent diffusivity \citep{Devlen_2013}.

The first flow was chosen for our study for its simplicity and relevance to
cosmology and astrophysics since it manifests the alpha effect, which is important
in galaxy disks \citep[e.g.][]{RevModPhys.74.775,annurev:/content/journals/10.1146/annurev-astro-071221-052807}.

To test the induction in the \textsc{Swift} code, the kinematic dynamo regime
was chosen. This is the regime where the velocity field is not affected by the
back-reaction from the magnetic field via the Lorentz force. This renders the
equations linear and allows different modes (sourced in the initial conditions)
to grow independently. On the code level we force the particles to obey at all
times the following velocity vectors:
\begin{equation}
    \textbf{v}_{a} = \textbf{v}_{\rm f}(\textbf{r}_{a}).
\end{equation}
where $\textbf{v}_{\rm f}$ is the forcing velocity field given below.

\subsubsection{Detailed setup}

For the velocity field of the flow, we use the same convention as
\cite{Rheinhardt_2014}:

\begin{align}
    v_{{\rm f},x} = \hspace{0.23cm} v_{\rm 0}\sin k_{\rm 0}x \cos k_{\rm 0}y \notag \\
    v_{{\rm f},y} = -v_0\cos k_{\rm 0}x \sin k_{\rm 0}y \notag \\
    v_{{\rm f},z} =\hspace{0.23cm} \omega_0 \sin k_{\rm 0}x \sin k_{\rm 0}y \notag
\end{align}
where $v_{{\rm f},i}$ are the components of the particle velocity, $x,y,z$ the
particle positions, and $v_{\rm 0}, \omega_{\rm 0}$ the magnitudes of the planar
and vertical velocities respectively.  In our configuration, we fix $k_{\rm 0} =
2\pi/L_{\rm box}=1$ and $\omega_{\rm 0}/\sqrt{2} = v_{\rm 0} = 1$ such that the
spread in velocities is normalised $v_{\rm rms} = 1$.

For the magnetic fields, we use three separate initial configurations:\\
First, \emph{random initial conditions (ICs)}: We generate a uniform distribution of vector
potentials $\textbf{A}$ with random length in range [0,$A_0$].
While initially the vector potential is not smooth, the magnetic
field is calculated through SPH version of the equation $\textbf{B} = {\rm
  curl}\textbf{A}$ (Eq. \ref{MF_from_VP}) is
spatially smooth and divergence-less form of $\vec{B}$ by construction.\\
Second, \emph{random ICs without bulk MF}: calculation of the magnetic field from random vector potential
may introduce a small constant field component.
Such components along the $z$ direction will not decay, and thus can influence
decaying modes (see Sec.~\ref{subsec:growth_rate_comparison} below).
Thus, the volume averaged magnetic field is subtracted. \\
Third, we employ a \emph{Beltrami-type field}, i.e. a field for which the
relation ${\rm curl} \textbf{B} = k \textbf{B} $ holds, which is also
divergence-less. More specifically, we use:
\begin{align}
  B_{x} &= B_0 (\sin k y + \sin k z), \notag \\
  B_{y} &= B_0 (\cos k x - \cos k z), \notag \\
  B_{z} &= B_0 (\sin k x + \cos k y). \label{Beltrami_field} 
\end{align}
The constants $B_0, A_0$ are the magnetic field and vector potential
normalization factors respectively.
Setting the magnetic field wavenumber to $k=1$ was enough to start
the mode with the largest growth rate in the range $\eta \in [0.1,0.2]$
(see Sec.~\ref{subsec:growth_rate_comparison} below).
We set the initial field $B_0$ or $A_0$ such that $B_{\rm rms}(t=0) = 10^{-4} B_{\rm eq}$, where equipartition estimate is:
\begin{equation}
  \epsilon_{\rm mag}\simeq \epsilon_{\rm kin}, \hspace{0.5cm}  \epsilon_{\rm mag}= \frac{B_{\rm eq}^2}{2\mu_0},  \hspace{0.5cm} \epsilon_{\rm kin}= \frac{v_{\rm rms}^2}{2},
\end{equation}
where $k=2\pi/L$ for our domain of size $L^3$.

As mentioned above, we make use of Eq.~(\ref{Rm_definition}) for the magnetic Reynolds
number definition.
For completeness, we chose a value of the internal energy, $u$, of order $1000$ such that the resulting flow remains sub-sonic at all times. Note that, in this setup, the sound speed only affects the time-step size (via the Courant condition) and the Dedener cleaning speed.


\subsubsection{Qualitative Roberts flow I results}

We study the Roberts flow I problem using $16^3$, $24^3$, $32^3$, $48^3$, $64^3$
and $128^3$ particles initially arranged in a glass-like (random) setup with
uniform density\footnote{A glass arrangement can be obtained by starting with
particles in a cubic grid lattice with small perturbations in lattice positions,
with no pressure gradients or velocities and by letting particles move under
numerical forces to a configuration where the kinetic energy does not change any
more.}.

To ensure that \textsc{Swift} correctly reproduces Roberts flows, we expect the magnetic field distribution, $\textbf{B}(x,y,z)$, across the simulation volume to be comparable to that obtained from other MHD codes. While the amplitude of the magnetic field in a growing mode increases exponentially with time, its normalized spatial distribution—given by $\textbf{B}(x,y,z,t)/B_{\text{rms}}(t)$ —should remain time-independent.

For code comparison, we use data from a Roberts flow simulation performed with the \textsc{Pencil Code}.
This simulation follows the setup described above, initialized with a
random vector potential at a physical resistivity around $\eta \simeq 0.18$.
Since Roberts flows exhibit translational symmetry along the $z$ direction,
a growing mode initialized with random conditions can manifest with an arbitrary shift along $z$.
To eliminate this ambiguity in the \textsc{Swift} simulations, we opted for a single-mode initial condition (Eq. \ref{Beltrami_field}), which ensures that the growing mode appears without an arbitrary vertical shift.

Figure~\ref{fig:RF1_field_configuration} compares the spatial configuration of the magnetic field components
between \textsc{Swift} and \textsc{Pencil Code} runs.
The upper row displays the magnetic field distribution in a $32^3$ - resolution \textsc{Pencil Code} simulation,
taken as a grid cell values at $z=0$.
The lower row presents results from a $32^3$ - resolution \textsc{Swift} run, sliced at $z = 0.05 \, L_{\text{box}}$.
In both cases, the plots show density maps of the individual magnetic field components, normalized by the root mean square magnetic field over the simulation volume. The visual agreement between the two codes indicates that \textsc{Swift}’s direct induction MHD implementation accurately reproduces the expected field configuration,
demonstrating consistency with the \textsc{Pencil Code} results.
We also verified that the vector potential implementation leads to similar
results.

Having verified that our SPMHD implementation agrees qualitatively with the
reference implementation in the \textsc{Pencil Code}, we move on to quantitative
measurements of the growth rate of the field.


\subsection{Growth rate comparison}
\label{subsec:growth_rate_comparison}

The instantaneous growth rate, $\gamma_i$ , is defined as:
\begin{equation}
    \gamma_{i} = \frac{d}{dt} {\rm ln}(B_{\rm rms}),
\label{instant_growth_rate}
\end{equation}
where $B_{\rm rms}$ is the root-mean-square average of the $B$-field in the
whole simulation volume. The instantaneous growth rate is averaged on the
interval where it is roughly constant. For our runs, specifically, we choose to
measure quantities at an interval $\delta t = 0.05$ in internal time units over
the range $t\in [30,70]$, where the growth rate is indeed well-behaved and
constant. The growth rate and fluctuations are then calculated as:
\begin{equation}
    \gamma = \langle \gamma_{i} \rangle_{\rm t}, \hspace{0.5cm} \delta\gamma = 2 \cdot \sqrt{\langle (\gamma_{i}-\gamma)^2 \rangle_{\rm t}}.
\end{equation}

\begin{figure}
    \centering \includegraphics[width=\linewidth]{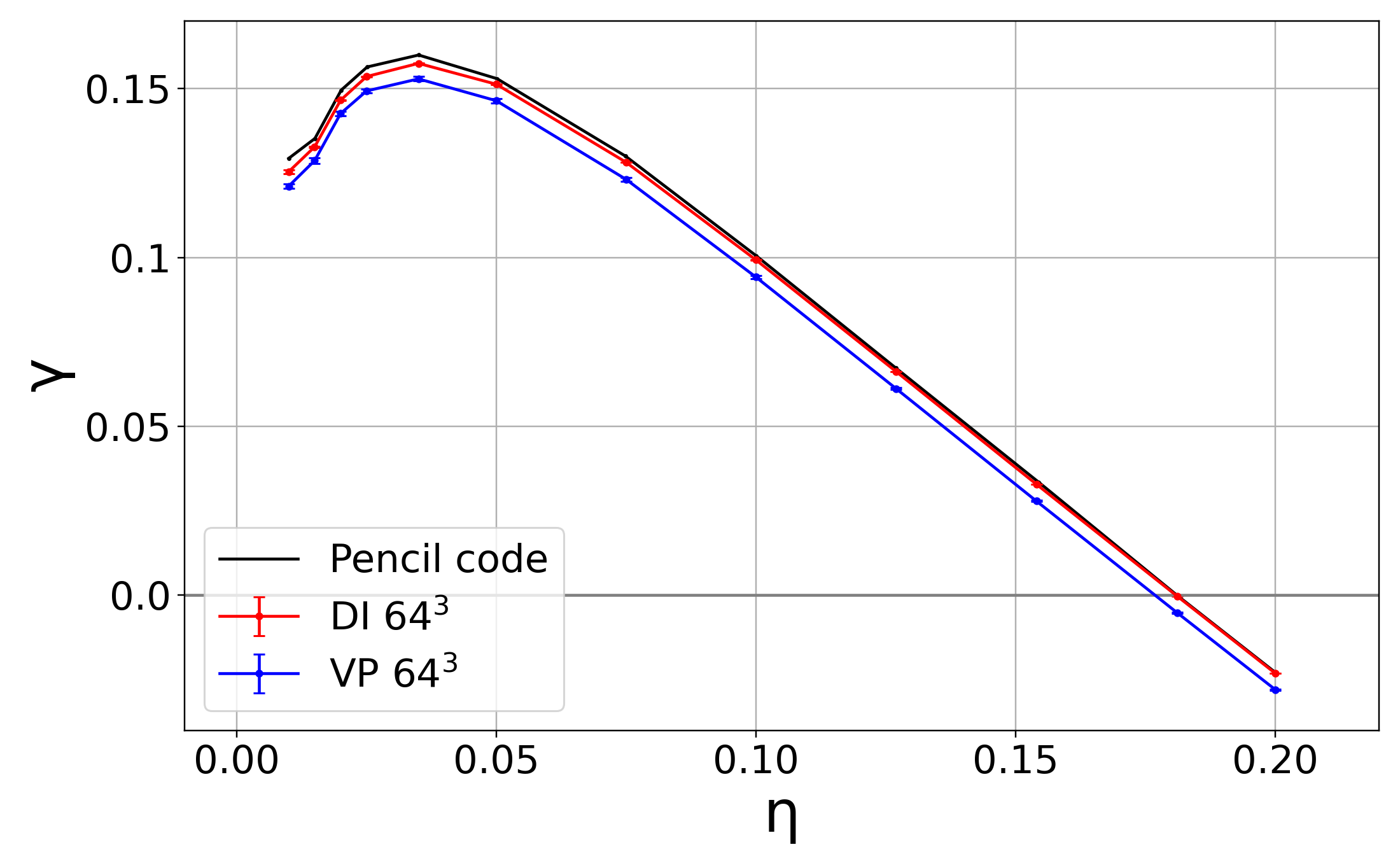}
    \includegraphics[width=\linewidth]{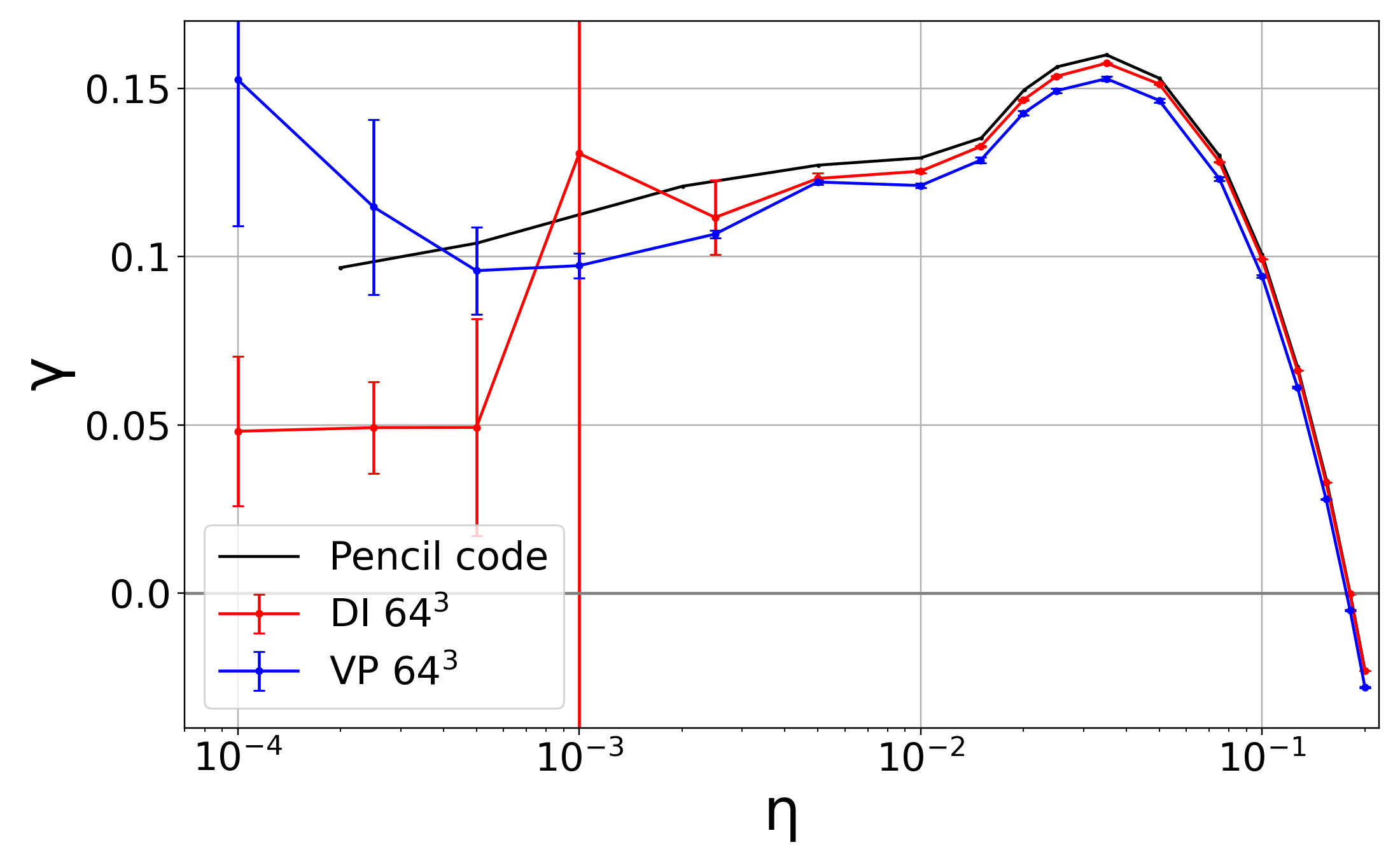}

    \caption{Top panel: The growth rate of the Roberts Flow 1 dynamo as a
    function of the physical Ohmic resistivity for the \textsc{Pencil} code
    with $32^3$ cells (black line) compared to both \textsc{Swift} MHD
    implementations using $64^3$ particles (red, blue). The bottom panel shows
    the same data but using a logarithmic axis for the resistivity to focus on
    the behaviour of the growth rates close to the ideal MHD limit
    ($\eta\rightarrow0$). Negative growth rates indicate
    decaying modes. The error bars represent the magnitude of numerical growth
    rate fluctuations. The direct induction (DI, red curve) implementation of
    SPMHD in \textsc{Swift} matches the \textsc{Pencil Code} results closely for
    almost all values of the resistivity (and thus Reynolds number); the
    cross-over point between growing and decaying modes, in particular, matches
    precisely. Deviations are only found close to the ideal MHD limit. The
    vector potential (VP, blue curve) implementation displays a systematically
    lower growth rate than the other models.
 } \label{fig:RF1_growth}
\end{figure}

In the Roberts flow setup, the excited magnetic field mode grows exponentially
with time and has a growth rate that depends directly the magnetic Reynolds
number. In our experiments, the velocity and wavelength of the flow are fixed
and the physical resistivity is varied in order to change $R_{\rm m}$. We report
the resulting dependence of the growth rate on the resistivity in
Fig.~\ref{fig:RF1_growth}.
The top panel depicts the growth rate as a function of resistivity near the dynamo onset, while the bottom panel illustrates its behavior in the ideal MHD limit $(\eta \to 0)$.
Note that the $x$-axis in the bottom panel is log-scaled for resistivity.
The black line represents results from \textsc{Pencil} code simulations at a $32^3$ resolution, whereas the red and blue lines correspond to the DI and VP MHD implementations in \textsc{Swift} code at $64^3$ resolution.
In the \textsc{Pencil Code}, random ICs were used for all data points.
For the \textsc{Swift} code, random ICs were applied for $\eta < 0.1$, while one-mode ICs were used for $\eta \sim [0.1,0.2]$.
The top panel shows growth rate as a function of resistivity near the dynamo onset.
The bottom panel illustrates how the growth rate behaves in the ideal MHD limit,
$\eta \rightarrow 0$ (note that for the bottom panel, the $x$ axis shows the log scaled resistivity).
Black line shows $32^3$ \textsc{Pencil}  code runs, red and blue show DI and VP MHD implementations in \textsc{Swift} code at $64^3$ resolution. The random ICs were used for \textsc{Pencil} code for all points, while for \textsc{Swift} code the random ICs were used for $\eta<0.1$ and one-mode ICs for $\eta \in [0.1,0.2]$.
Both MHD flavors in \textsc{Swift} exhibit slightly lower growth rates compared to the reference results from the \textsc{Pencil} code. However, both SPMHD implementations accurately capture the overall relationship between $\gamma$ and $\eta$ for high resistivity. The VP scheme underestimates growth rates relative to the DI scheme. One possible explanation is that, in the version of the code used for this study, parameters in the gauge equation were not as well optimized for the VP implementation as they were for the cleaning in DI.

The errors in Eq.~(\ref{MF_from_VP}) probably do not influence
the measurement of magnetic field growth since the spacial distribution of vector potential is time independent for a growing mode, such that $\textbf{A}(x,y,z,t) = f(t)\textbf{A}(x,y,z)$.

The lower
resolution runs, $32^3$ and $16^3$ exhibit the same behavior, but the deviation
from the \textsc{Pencil} code is larger for smaller resolutions.\\


The limit of small physical resistivity is crucial for the future \textsc{Swift}
applications to cosmology and astrophysics since it translates to large ($R_m
\gtrsim 10 \sim 100 $) magnetic Reynolds number. In the Roberts flow I case, when decreasing
the resistivity down to $\eta\simeq 3\cdot 10^{-3}$ (i.e $R_{\rm m}\simeq 300$),
the growth rate deviation of the SPMHD implementations from the \textsc{Pencil}
code increases (bottom panel of Fig.~\ref{fig:RF1_growth}). For our runs, the
growth rate is not the same and shows some resolution-dependent
fluctuations. These fluctuations are illustrated as error bars on the figure
where we used 2 standard deviations as the size of the bar. For growing mode
solutions, the relative magnitude of the growth rate uncertainty is at the level
of $10^{-2}$ but increases for lower resistivity and for lower resolution. At
$\eta \simeq 10^{-3}$--$10^{-4} $ (or $R_{\rm m} = 1000$--$10000$) the
growth rate fluctuations become of the same order as the growth rate itself.


\begin{figure}
    \centering
    \includegraphics[width=\linewidth]{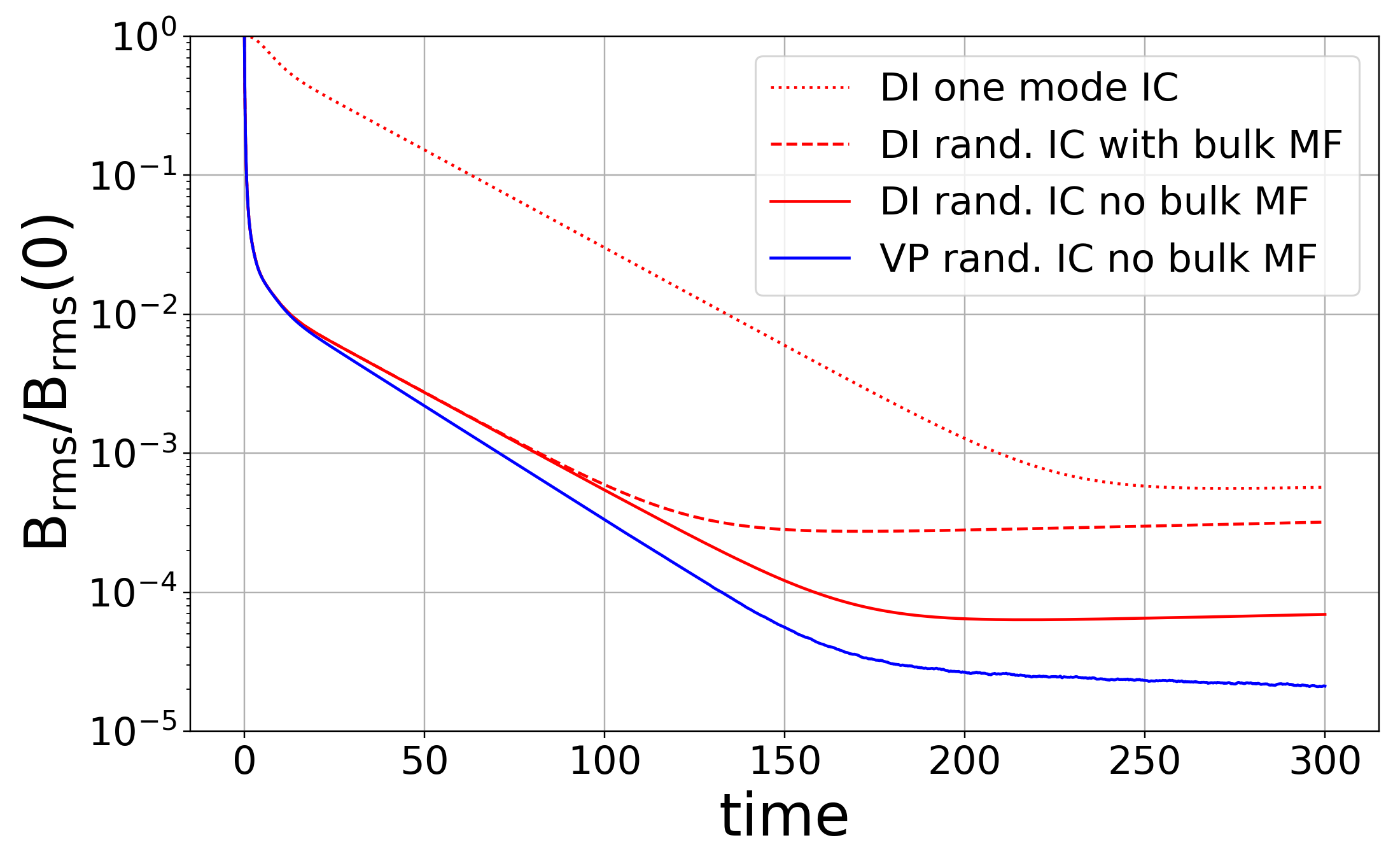}
    \caption{Magnetic
    field time evolution for decaying mode with different ICs: with one specific
    mode excited (red, dotted line), with random initial magnetic without bulk
    magnetic field subtraction (red, dashed line), with random magnetic field
    with bulk field subtracted (solid red - DI, solid blue - VP). The
    single-mode ICs consist of a Beltrami field with wavelength $\lambda =
    L_{\rm box}$. For all of the runs the correct decaying mode appears
    initially, but for DI is followed by other, slowly growing mode and slowly
    decaying modes. The runs indicate that longevity of the decaying mode can be
    improved by bulk field subtraction. However, the longer time behavior is not
    connected to the bulk field subtraction. Since one-mode ICs provide the
    longest existence time for the decaying mode, this initial field
    configuration was chosen to study the dynamo onset resistivity in what
    follows. }  
    \label{fig:RF1_growth_ic_dependance}
\end{figure}

Focusing now on the last two points from the right, (i.e. $\eta = 0.1811, 0.2$),
from Fig.~\ref{fig:RF1_growth}, the decaying mode forms, as expected. Our
experiments show, however, that if the random ICs (rather than the Beltrami
field) are used the negative growth rate does not become constant with time and
slowly tends to zero with the magnetic field reaching some plateau and only
slowly changing with time. This behaviour is highlighted on
Fig.~\ref{fig:RF1_growth_ic_dependance} where we tested both MHD implementations
and and three types of initial conditions for the DI case:
\begin{itemize}
\item random magnetic field with bulk field subtracted,
\item random magnetic field without bulk field subtraction,
\item single-mode ICs, i.e. Beltrami field.
\end{itemize}
For all schemes and ICs, the decay stops. But with
one-mode ICs the decaying modes exist for a more extended period of time, which
helps the precise measurement of the growth or decay rate. We thus use such
single-mode ICs in the following precision tests, in range $\eta \sim [0.1,0.2]$. For small resistivity, $\eta \rightarrow 0 $, the modes with $k<1$ become important \citep[e.g.][]{10.1098/rsta.1972.0015}. For this reason we used random ICs in $\eta <0.1$, because they contain a spectrum of modes, with wavelength spanning from resolution scale up to the simulation boxsize. 

\subsection{Transition from growing to decaying regime}

\begin{figure}
    \centering
    \includegraphics[width=1\linewidth]{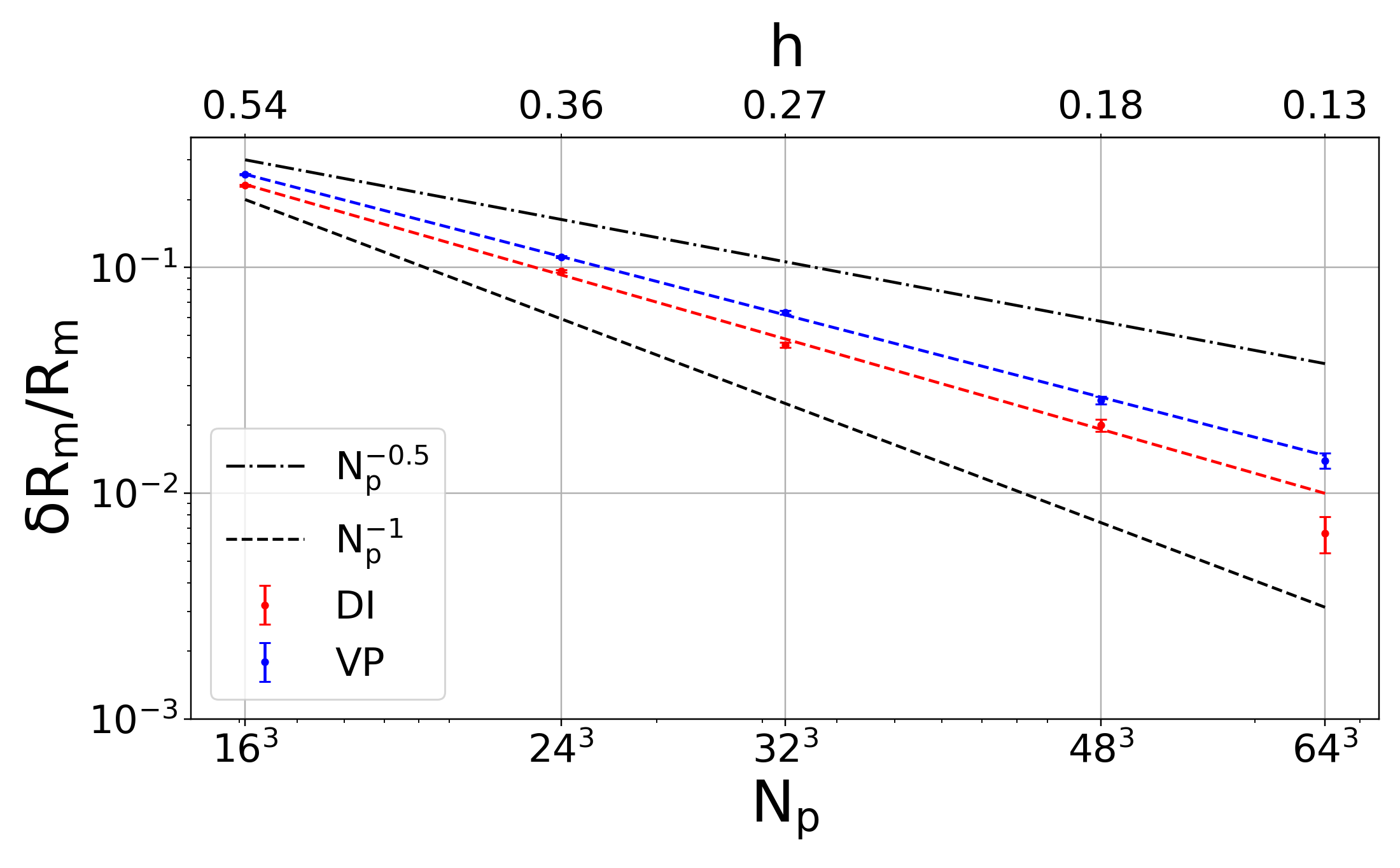}
    \caption{Relative deviation of the critical Reynolds magnetic number of the
             Roberts flow I problem from the one extracted from our
             highest-resolution ($N=128^3$ particles) as a function of the
             number of particles (bottom axis), or smoothing length (top axis)
             in a simulation of size $L_{\rm box}=2\pi$. The black lines
             indicate slopes of $N^{-0.5}$ and $N^{-1}$. Both MHD
             implementations in \textsc{Swift} converge with increasing
             resolution at a relatively similar rate. }
    \label{fig:RF1_convergence_log}                
\end{figure}

An important aspect is the dynamo onset resistivity and the corresponding
$R_{\rm m}$. This is the critical resistivity at which the behaviour transition
from a growing to a decaying magnetic field. If captured incorrectly by an
implementation, kinetic dynamos can appear under incorrect conditions. In the
case of more complex simulations such as cosmological MHD simulations this can
result too much or too little magnetic field growth and can be harder to
identify than for less complex systems. The cosmological simulations can also
have some resolved and under resolved flows, and the resolution dependence of
critical magnetic Reynolds number $R_{\rm m}^{\rm crit}$ in the Roberts Flow I
test can thus give us insight to the dependence of the dynamo strength on the
flow resolution in different regions of cosmological
simulations. \cite{Tilgner_2008} measured the transition from decaying to
growing modes in the Roberts flow I test to occur around $R_{\rm m}^{\rm
crit}\simeq 5.52$, which corresponds to $\eta = 0.1811$ for the setup. The
results of the simulations are resolution dependent, but we should expect them
to converge to the correct $R_{\rm m}^{\rm crit}$. To validate the
implementations and obtain their convergence rates, we started by running a
series of additional simulations using $128^3$ particles around the expected
critical resistivity. We then measure the growth rate and interpolate between
the measured values to get $R_{\rm m}^{\rm crit}$. Using this procedure, we
find:
\begin{align*}
  R_{\rm m}^{\rm crit} &= 5.487 \pm 0.005\qquad{\rm (DI~implementation)}, \\
  R_{\rm m}^{\rm crit} &= 5.571 \pm 0.004\qquad{\rm (VP~implementation)},     
\end{align*}
in reasonable agreement with published values. 

To study the convergence rate, we repeat the same exercise but using simulations
with $16^3$, $24^3$, $32^3$, $48^3$, and $64^3$ particles. We then measure the
difference between the critical resistivity obtained at a given resolution and
the value extracted from our highest resolution run (reported above). This
difference is shown in Fig.~\ref{fig:RF1_convergence_log} as a function of the
particle number (bottom axis) or, equivalently, as a function of the SPH
smoothing scale, i.e. the spatial resolution (top axis). As can be seen, both
schemes do converge to the solution though at slightly different rates. We
additionally measure the convergence rate as a function of particle number by
fitting a power law to the data on the figure. We find slopes of:
\begin{align*}
  \alpha &= -0.762 \pm 0.023 \qquad{\rm (DI~implementation)}, \\
  \alpha &= -0.690 \pm 0.014 \qquad{\rm (VP~implementation)},
\end{align*}
i.e. a convergence rate close to second order in the spatial resolution ($\delta
R_{\rm m} \propto h^2$).

\subsection{Impact of resolution and resistivity}

\begin{figure*}
    \centering
     \includegraphics[width=0.24\linewidth]{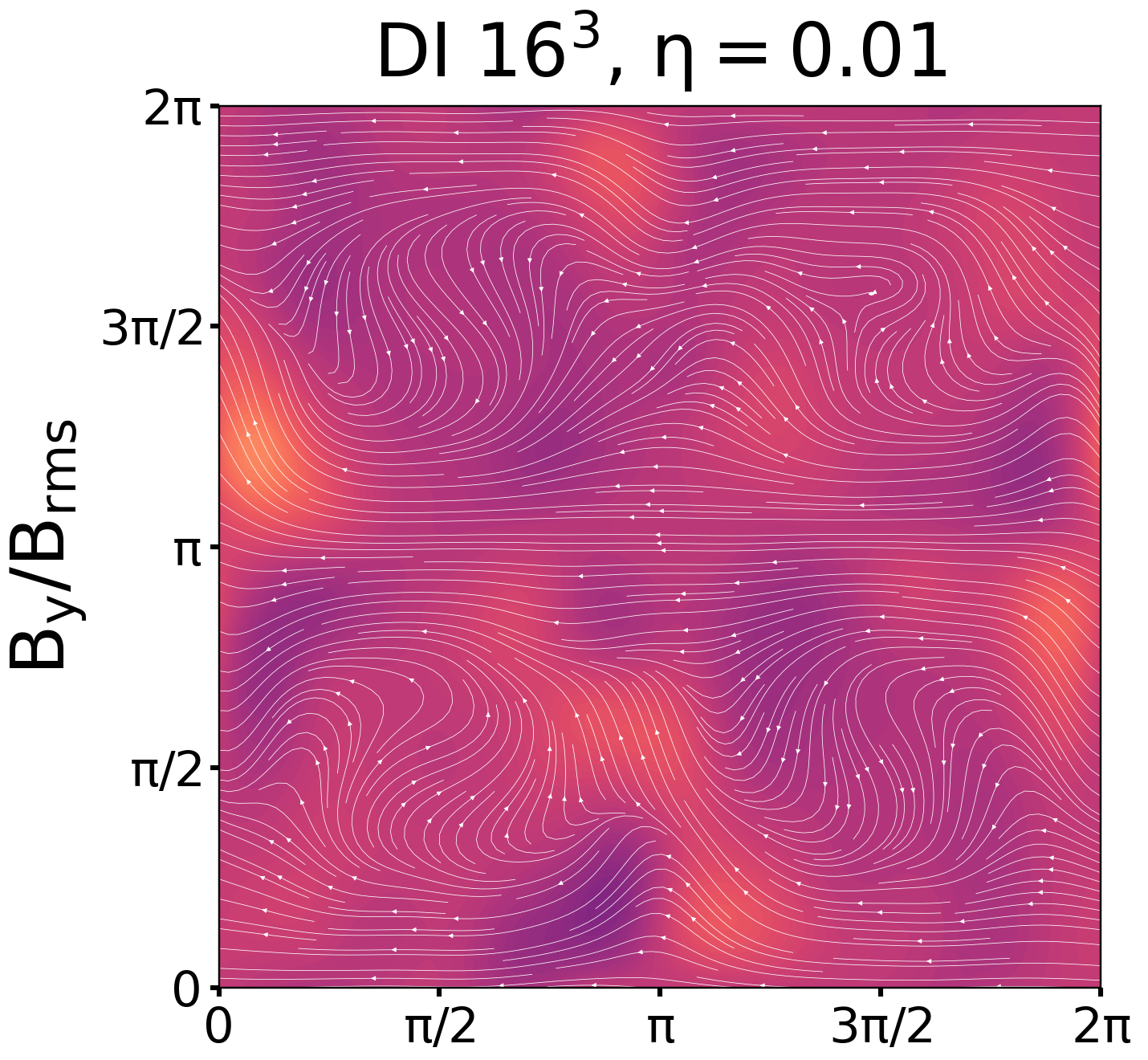}
     \includegraphics[width=0.24\linewidth]{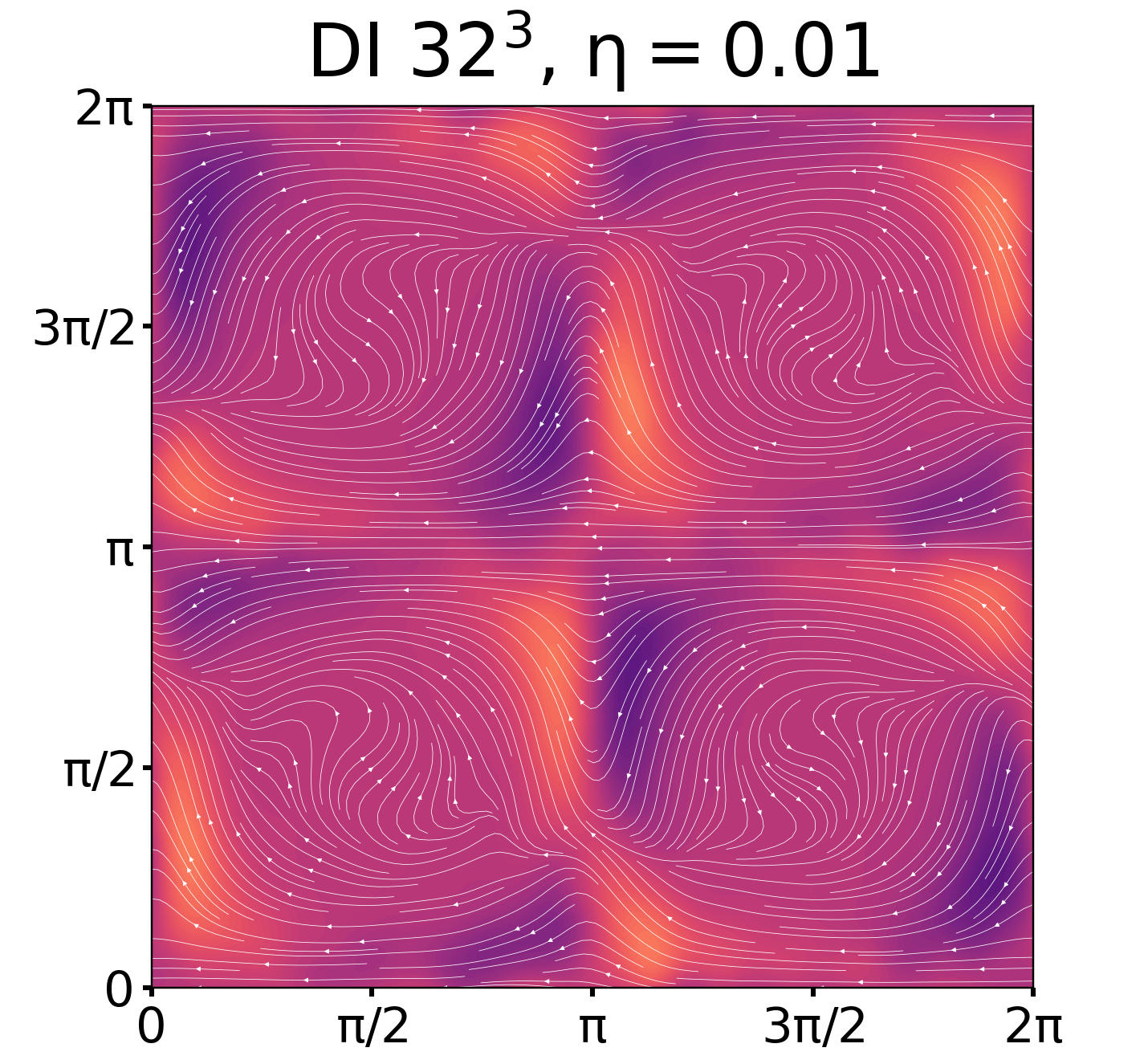}
      \includegraphics[width=0.24\linewidth]{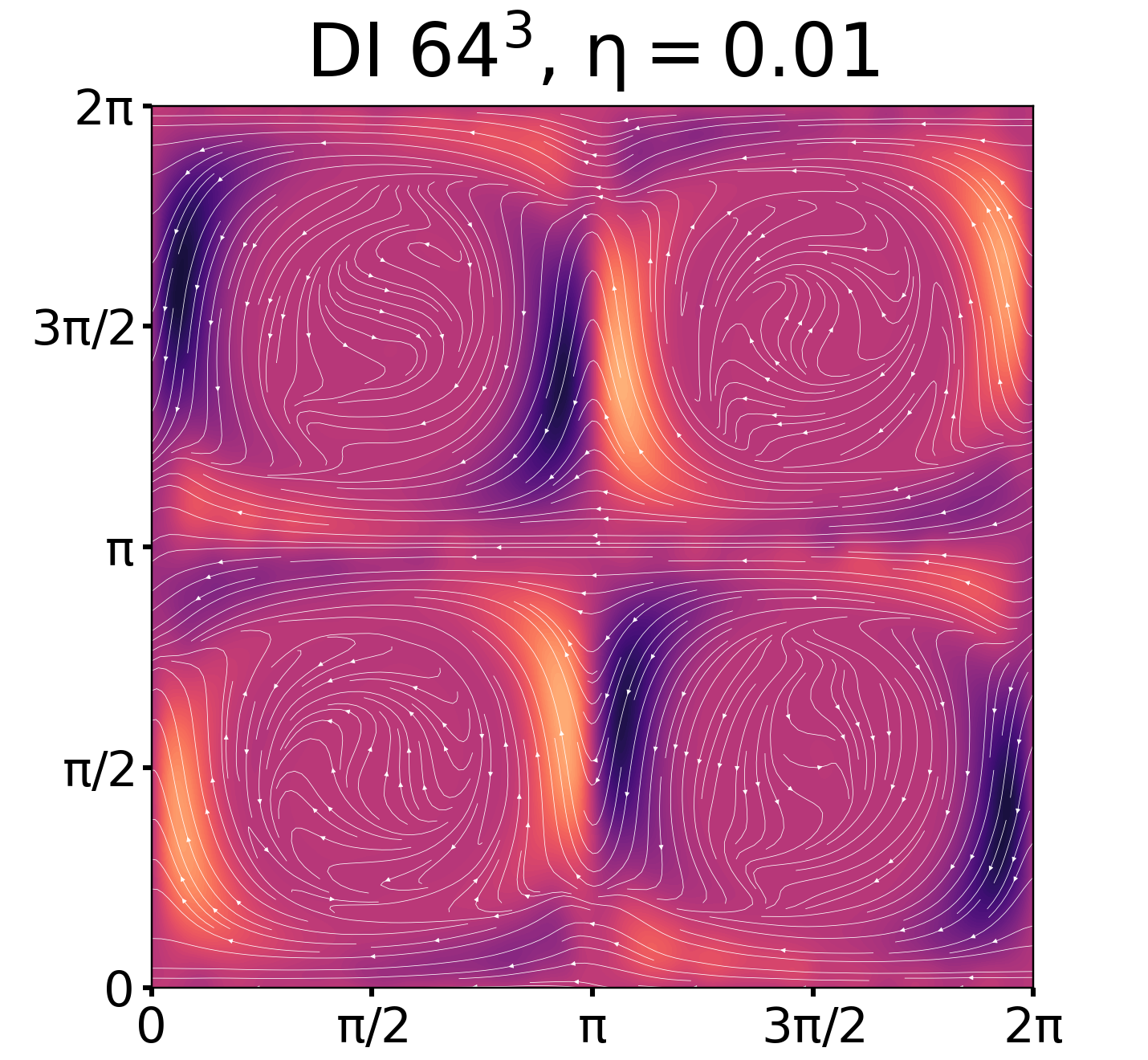}
      \vline
    \includegraphics[width=0.24\linewidth]{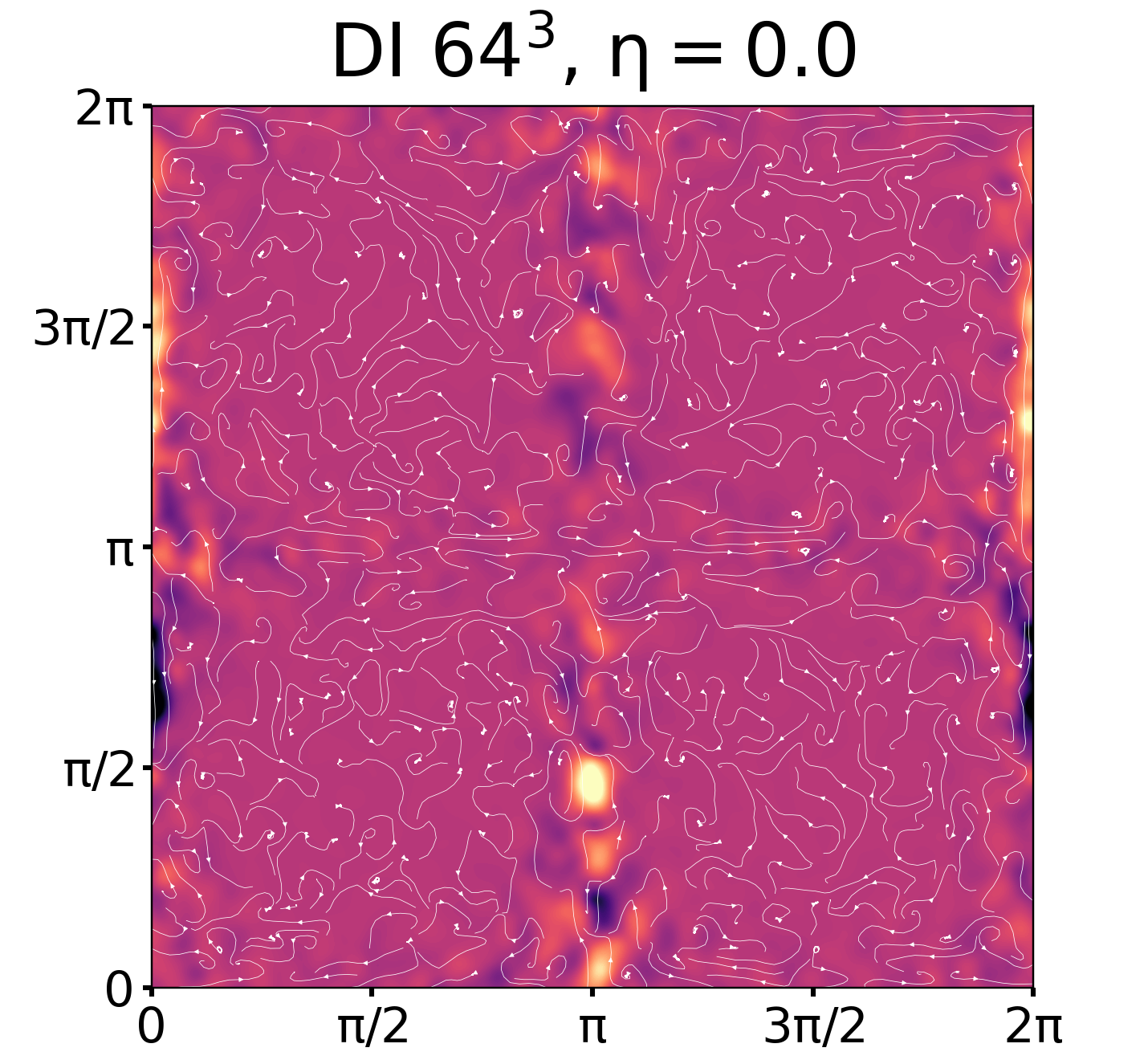}
    \caption{Colormaps for $B_{y}/B_{\rm rms}$ at $t=50$ in XY plane for
      Roberts Flow 1 runs at $16^3$ (left), $32^3$ (center left), $64^3$ (center
      right) resolution with $\eta=0.01$ and $\eta=0$ run at $64^3$ (right). For
      low resolution runs magnetic field pattern appears more distorted than at
      higher resolution. As the resolution is increased more thin magnetic field
      features become visible. In the limit of zero resistivity, the magnetic
      field pattern becomes unresolved for any resolution.}
    \label{fig:RF1_overwinding}
\end{figure*}

Magnetic field pattern develops thinner features as the resistivity
decreases. At some threshold resistivity and resolution, distortions of the
magnetic patterns appear. However, for higher resolution, at fixed $\eta$, the
distortions are absent, indicating the connection of magnetic field pattern
breakdown in the ideal MHD limit and resolution. The process of increasing
resolution at a fixed resistivity is visually illustrated in the left three
panels of Fig.~\ref{fig:RF1_overwinding}.
The figure presents density maps of magnetic field slices, depicting $B_{y}(x,y,z,t) / B_{\rm rms}(t)$ in the $xy$-plane for resolutions of $16^3$, $32^3$, and $64^3$ at $\eta = 0.01$, along with a $64^3$ run without resistivity. In all cases, random initial conditions were used, and the slice heights were adjusted to capture the same feature. The colormaps represent the magnitude of the magnetic field component, using the same color scale as in Figure \ref{fig:RF1_field_configuration}. White streamlines indicate magnetic field lines in the $xy$-plane.

As the resolution increases from left to right, initially unresolved structures reveal more detail, reducing blob-like distortions. However, in the zero-resistivity case, the pattern remains distorted at all resolutions, with elevated magnetic field regions appearing along the vortex boundaries ($x = \pi n$  and $y = \pi m$, where $n, m$ are integers), with a characteristic size on the order of the resolution scale (Figure \ref{fig:RF1_overwinding}, rightmost plot). The growth of magnetic distortions at the resolution scale in the absence of resistivity is not unique to the \textsc{Swift} DI or VP implementations and has also been observed in other MHD dynamo simulations \citep[e.g.][]{10.1111/j.1365-2966.2009.15640.x}.


\subsection{Influence of spurious magnetic field divergence}\label{divB}

\begin{figure}
    \centering
    \includegraphics[width=\linewidth]{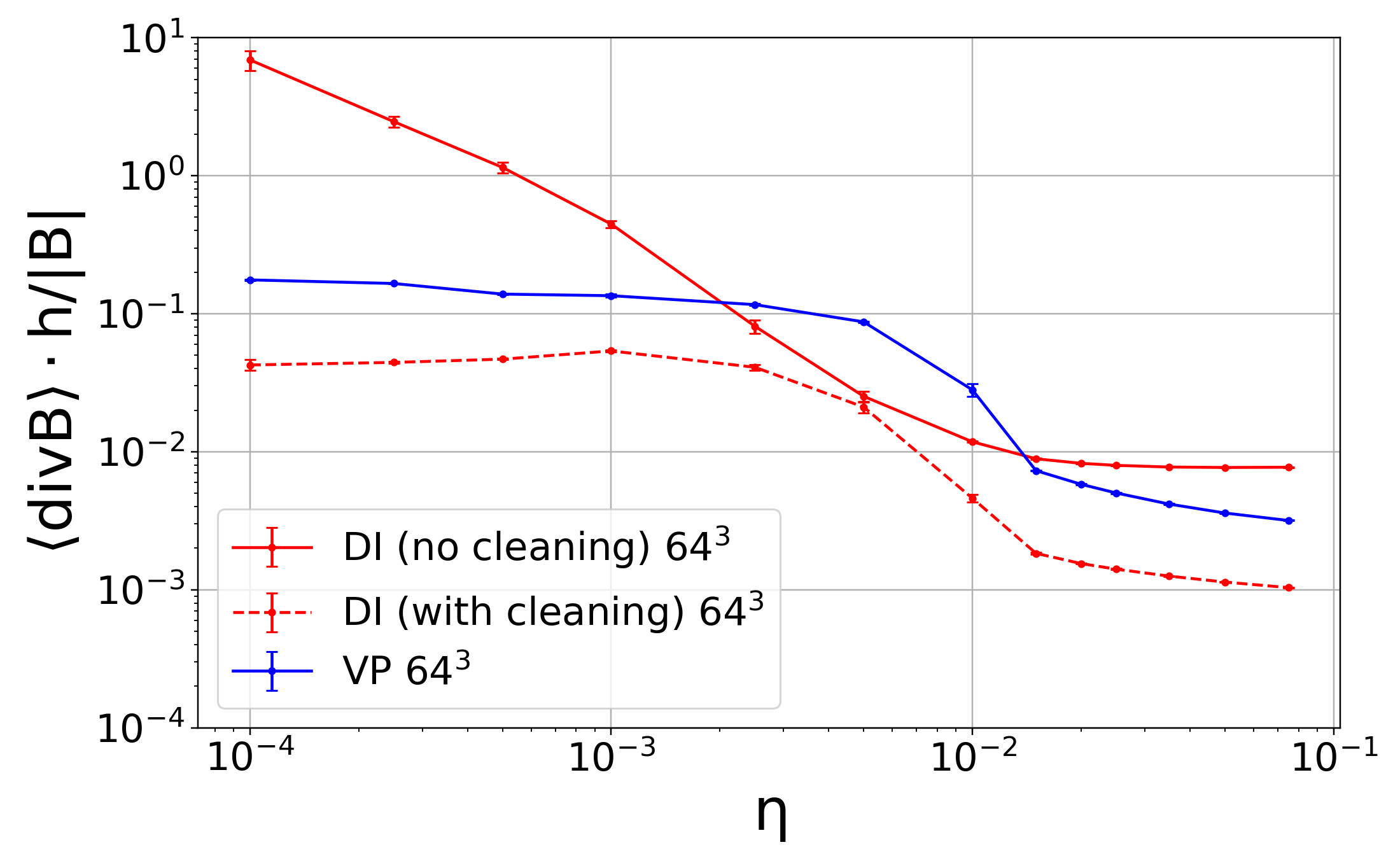}
    \caption{Magnitude of mean volume and time averaged divergence error vs
      physical resistivity for $64^3$ runs for \textsc{Swift} MHD
      implementations. The divergence errors here for both DI and VP are the
      measure of accuracy of the produced B field. For DI the divergence is also
      entering equations of motion. The error experiences order of magnitude
      jump around $\eta\sim 3\cdot 10^{-3}$ for both MHD
      implementations in the same region where large growth rate deviations from
      \textsc{Pencil} code appear. For direct induction scheme the Dedner
      cleaning helps to keep divergence errors where otherwise they will be big (red solid and dashed lines). }
    \label{fig:RF1_divB_err}
\end{figure}

Physical magnetic field should maintain the solenoidality condition (${\rm
div} \textbf{B} = 0$). Probing the level of spurious divergence can help indicate
how much the simulation results can be trusted. The velocity field forcing term
used for the Roberts flow test decouples the force equation from the induction
one. As such, the only way spurious divergence can affect the magnetic field
evolution is through the induction equation. Ideally, in the Eulerian frame, the induction term ${\rm curl}[\textbf{v} \times \textbf{B}]$ should not generate any divergence. However, due to inaccuracies in the SPH operator, the induction term may inadvertently introduce divergence. The presence of a monopole component in the magnetic field, $\textbf{B}_{\rm mon}$, can then act as a source for physical fields through the term ${\rm curl}[\textbf{v} \times \textbf{B}_{\rm mon}]$, potentially altering the magnetic field growth rate.

As an initial check, we monitored the mean divergence error throughout our runs. The divergence was evaluated once the growing mode was established, using a time-averaged measurement over the interval $t \in [30,70]$ and with volume-averaging over the entire simulation domain. The results are reported as a function of
resistivity on Fig.~\ref{fig:RF1_divB_err}.
The figure presents the divergence error, $R_0$, without noise cancellation for $64^3$ DI and VP runs. Now we also include the direct induction runs with measures to
clean the divergence (Sec.~\ref{ssec:ODI}).
For resistivity values above $\eta\simeq 3 \times 10^{-3}$, magnetic field is well behaved, the error remains below $10\%$ in all cases. However, for smaller resistivity, the errors increase significantly. In the VP runs, the error saturates at approximately $20\%$. In DI runs without cleaning, the error continues to grow as resistivity decreases until a significant portion of the field consists of monopole component. When Dedner cleaning is enabled, the errors are significantly reduced to an acceptable level of about $5\%$, and exhibit behavior more similar to VP. The DI implementation with cleaning results less errors than VP over whole resistivity range. 

\begin{figure*}
    \centering
    \includegraphics[width=0.22\linewidth]{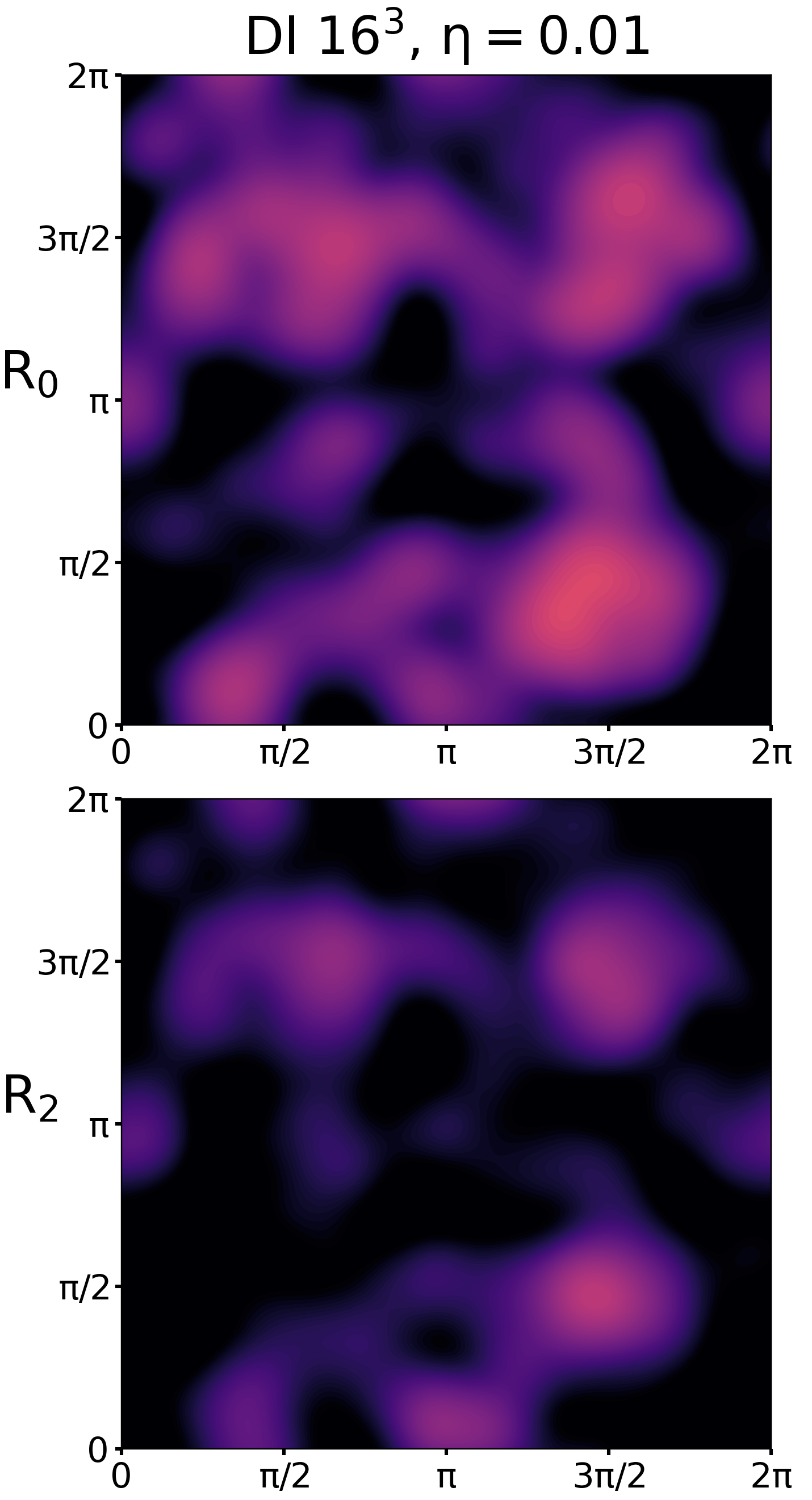}
     \includegraphics[width=0.22\linewidth]{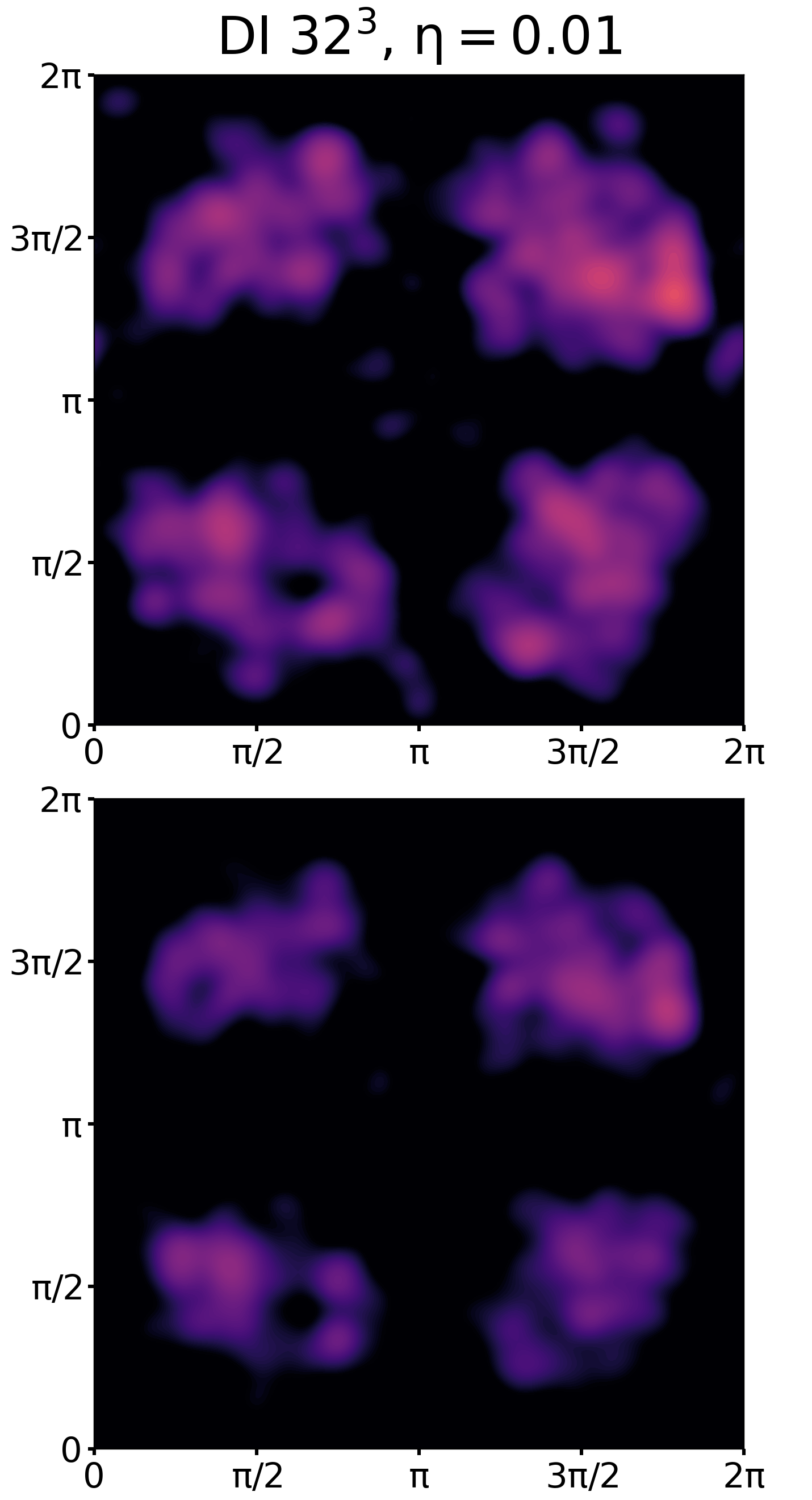}
      \includegraphics[width=0.22\linewidth]{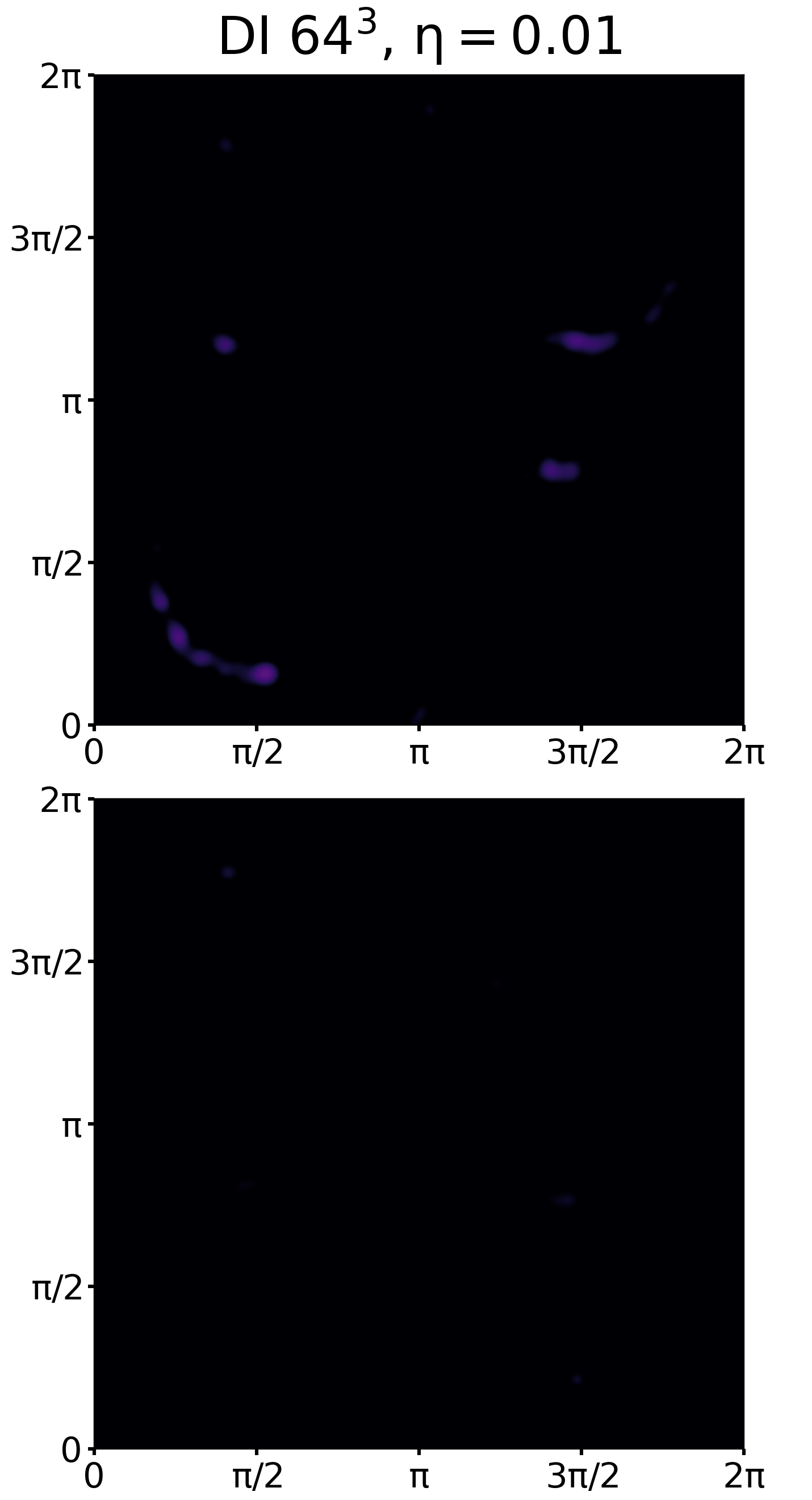}
      \vline
      \includegraphics[width=0.22\linewidth]{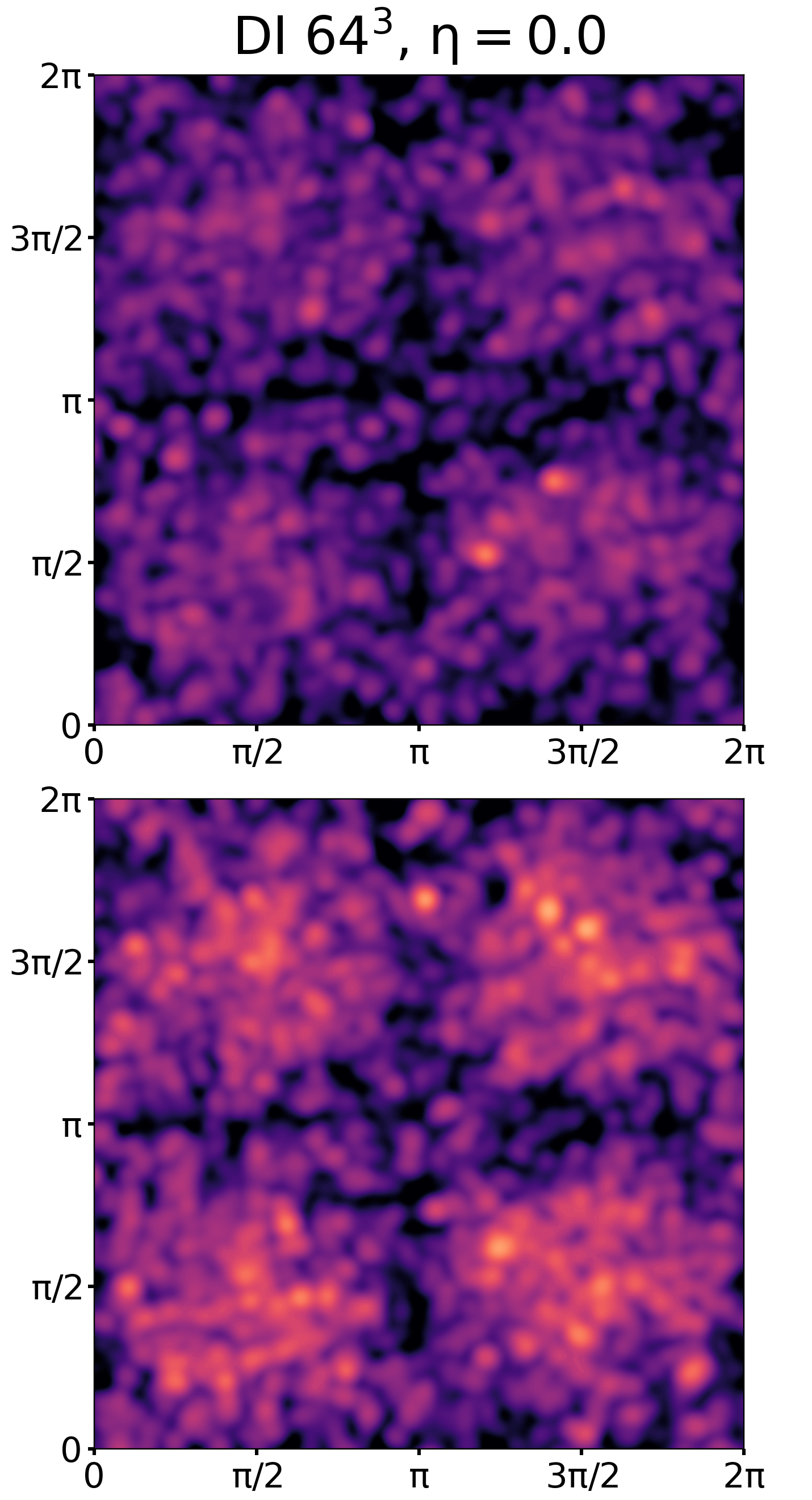}
      \includegraphics[width=0.0385\linewidth]{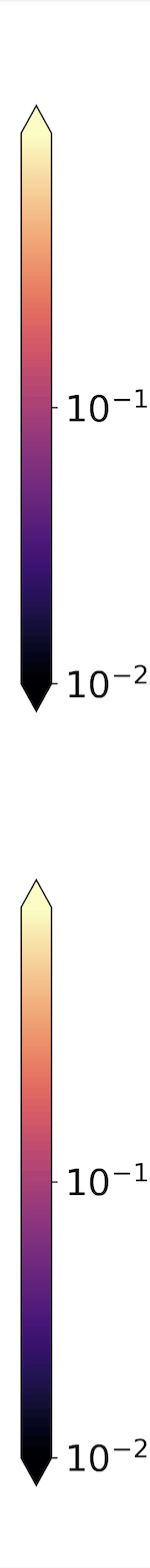}
      \caption{Density maps for dimensionless divergence error metrics
        $R_0={\rm div}\textbf{B} \cdot h/|\textbf{B} |$ (top) and
        $R_2= {\rm div}\textbf{B} / |{\rm curl}\textbf{B}|$ (bottom) with additional
        cuts to reduce SPH noise. The Roberts Flow I runs were performed with
        $\eta=0.01$ for 3 resolutions: $16^3$ (top column), $32^3$ (upper middle
        column), $64^3$ (lower column) particles , and an additional single run
        with $\eta=0$ at $64^3$ resolution (right-most panel). Regions with
        $R_{i}\gtrsim0.1$ indicate large divergence errors.  The divergence
        errors are elevated mostly inside vortices. Resolution increase leads to
        decrease of divergence error metrics. Both $R_0$ and $R_2$ maps are
        similar in general however have some small differences at resolution
        scale. Divergence errors are small even in the ideal MHD limit. However,
        the $R_2$ metric shows more elevated values in this regime.}
    \label{fig:RF1_error_metrics}
\end{figure*}

In addition to monitoring global averages, it is also instructive to examine the spatial distribution of divergence errors. We used two error metrics introduced in Sec.\ref{ssec:error_metrics}: $R_0$ and $R_2$. Colormaps of these error metrics for Roberts Flow I at $\eta = 0.01$ (where pattern destruction occurs for $N=16^3$ particles; see below) are shown in Fig.\ref{fig:RF1_error_metrics} for three different resolutions, along with an additional run in the ideal MHD limit. All runs were performed using the DI implementation with divergence cleaning enabled. These maps correspond to the same setup and time as those in Fig.~\ref{fig:RF1_overwinding}.

The mean errors remain small across the simulation volume for all resolutions but are more pronounced inside vortices, that correlates with large shear in the velocity of the Roberts flow. The $R_2$ metric highlights similar regions as $R_0$ but tends to yield slightly lower values. In the ideal MHD case (i.e., zero resistivity), the $R_0$ metric remains mostly below $10\%$, whereas $R_2$ exceeds $10\%$. This indicates that the two error metrics are not entirely equivalent, and $R_0$ alone does not fully capture all unphysical magnetic field.


\begin{figure}
    \centering
     \includegraphics[width=1\linewidth]{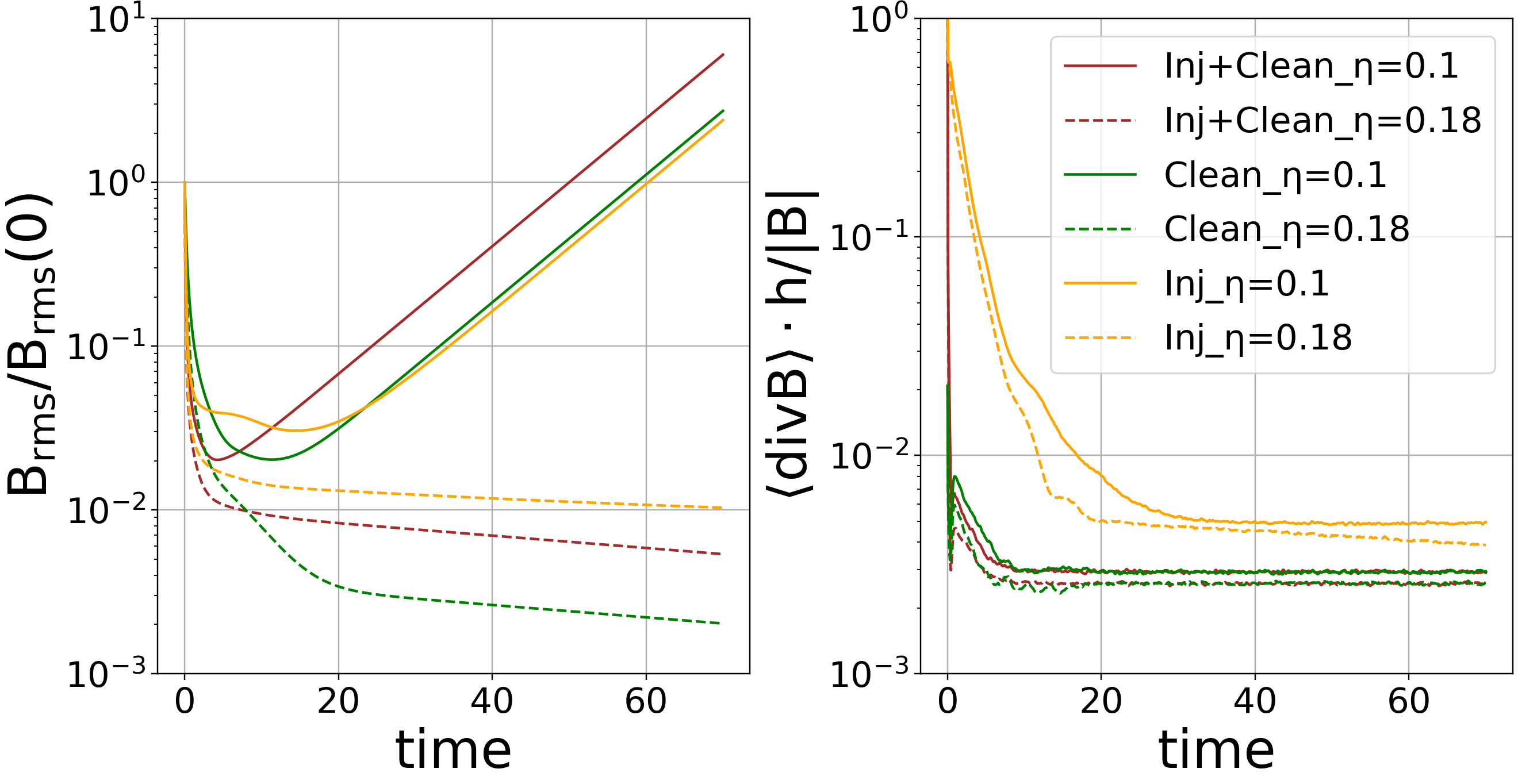}
    
     \caption{Evolution of the magnetic fields (left) and divergence errors
       (left) on a Roberts flow I test case with $32^3$ particles using the
       direct induction (DI) MHD implementation in \text{Swift}.  We perform
       runs with a large initial injected divergence (Inj) and the Dedner cleaning (Clean) term
       switched on (brown) or off (orange), and with clean ICs with the Dedner
       term still on (green). We perform the tests for growing (solid line) and
       decaying (dashed line) modes. The divergence errors decrease quickly and
       don't affect the magnetic field growth rates. The addition of the Dedner
       cleaning terms helps to decrease the error before the divergence gets
       cleaned by resistivity.}
    \label{fig:Divergence_injection}
\end{figure}

The Roberts flow I is an idealized setup where the divergence can remain low both in space and time, and particle distributions, along with other quantities, are smooth. However, in real astrophysical and cosmological simulations, strong density, magnetic field, and velocity contrasts exist, and the particle distribution is highly non-uniform. Additionally, processes such as energy injection from subgrid models and particle removal can introduce sudden spikes in divergence. Therefore, it is crucial to investigate the impact of large divergence errors on the growth rate, evaluate the effectiveness and determine the main source of divergence cleaning.

To assess the cleaning performance in a dynamo setting, we conducted additional runs in which a large divergence error was deliberately introduced in the initial conditions. This was achieved by generating random initial conditions similar to ones described in Sec.~\ref{ssec:RF1}, but with randomly oriented $\textbf{B}$ field vectors instead of random $\textbf{A}$, thereby ensuring a significant monopole component in the magnetic field.


Figure~\ref{fig:Divergence_injection} illustrates the evolution of the magnetic field and divergence error over time for the DI implementation. Initially, divergence cleaning was disabled (yellow lines), and we observed that physical resistivity alone was sufficient to dissipate the large initial divergence error. However, when divergence cleaning was enabled, the errors were removed much more rapidly (brown lines). Ultimately, the divergence error saturated at the same level as in runs with divergence-free (random ICs from Sec. \ref{ssec:RF1}) initial conditions (green lines).

Thus, while some error reduction occurs naturally due to the resistivity term, the Dedner cleaning terms provide a significantly more effective correction.

The growth rates of growing modes (solid lines) remained unaffected by the divergence cleaning. However, for decaying modes (dashed lines), the run with resistivity alone exhibited a slightly slower decay compared to the runs with Dedner cleaning, suggesting the presence of a small additional diffusion effect from the cleaning process.
  


\subsection{The ABC flow}\label{ABC}
Before closing this section on classical dyanmo tests, we briefly report the
results of our experimentations with the more complex Arnold–Beltrami–Childress
\citep[ABC;][]{Arnold_2014,Childress_1970} flow.

For the previously considered Roberts flow I, the modes were growing exponentially while for the ABC flow they can manifest oscillatory behavior. In addition, the ABC flow represents a more complex flow where $\textbf{v}$ has a dependency in all 3 spatial directions. Properties such as the growth rates and the oscillation frequency of the ABC flowsi
were studied numerically \citep{Galloway01071986, Bouya2012RevisitingTA, Brandenburg_2020}.




Our setup used for the ABC flow uses the same mechanism for particle flow
forcing $\textbf{v}_{\rm f}$ as the one used for the Roberts Flow I.
The velocity field for the flow follows \cite{Bouya2012RevisitingTA}:
\begin{align}
    v_{{\rm f},x} = A\sin k_0 z+C\cos k_0 y \notag \\
    v_{{\rm f},y} = B \sin k_0 x + A \cos k_0 z \notag \\
    v_{{\rm f},z} = C \sin k_0 y + B \cos k_0 x
    \label{ABC_flow}
\end{align} 
where $v_{{\rm f},i}$ are the components of the forcing velocity and $x,y,z$ the particle
positions. We choose $A,B,C = 1/\sqrt{3}$ such that $v_{\rm rms} =1$, $k_0 =
2\pi/L_{\rm box}=1$. The random ICs from Sec. \ref{ssec:RF1} were used for the magnetic field. 

The most commonly used parameters for the symmetric ABC flow are $A = B = C = 1$, resulting in a reference root-mean-square velocity of $v_{\rm rms}^{\rm ref} = \sqrt{3}$. To facilitate a clearer comparison of growth rates with the Roberts Flow I runs, we set $v_{\rm rms} = 1$ in our \textsc{Swift} tests. This ensures that the system size and typical velocity match those of the Roberts Flow I, with the only difference being the flow geometry.

In this setup, comparing growth rates, frequencies, and Reynolds numbers with the reference requires applying a time and coordinate transformation. Notably, \cite{Bouya2012RevisitingTA} do not provide an exact definition of $R_{\rm m}$ as in Eq.~\ref{Rm_definition}, but instead define $R_{\rm m}$ through the induction equation:
\begin{equation}
    \frac{\partial\textbf{B}}{\partial t} = {\rm curl} [\textbf{v}\times \textbf{B}] + \frac{1}{R_{\rm m}} \Delta \textbf{B}.
    \label{Rm_def_through_MHD}
\end{equation}
To match our convention, we performed a time translation  of this equation: $t
= t^{\rm ref}\cdot \sqrt{3}$. As both the reference and \textsc{Swift} simulations have
$L = L^{\rm ref} = 2\pi$ for the flow periodicity, this leads to the velocity
relation $ v = v^{\rm ref}/\sqrt{3}$. Similarly, the Growth rates and frequencies relate as
inverse times: $\gamma,\omega = \gamma,\omega^{\rm ref}/\sqrt{3}$. Since in both
cases same the MHD equation are solved the magnetic Reynolds number relate as $R_{\rm
  m} = R_{\rm m}^{\rm ref} \cdot \sqrt{3}$

To measure growth rates of oscillatory modes in \textsc{Swift}, we track the peaks of the root-mean-square (RMS) magnetic field over the simulation volume as a function of time. In $\ln(B_{\rm rms})$ vs. $t$ space, these peaks exhibit a linear trend. We determine the growth rate and its associated error by performing a linear fit to $\ln(B_{\rm rms})$ at the peak points.

For frequency measurements, we employ two methods: 1) Measuring the mean time interval between successive peaks.; 2) Performing a Fourier transform of the instantaneous growth rate. The first method provides high accuracy when a single dominant growing mode is present. However, it becomes less reliable when multiple oscillatory modes with similar amplitudes and growth rates coexist, as seen in mode crossing within the ABC flow. The second method’s accuracy is constrained by the total simulation time.
To ensure robustness, we compute the frequency using both methods and report the value with the smallest error.
Note that one oscillation period in $B_{\rm rms}(t) $ corresponds to a magnetic field (MF) direction flip. Therefore, the full flip oscillation period is twice that value.

We do not measure growth rates and frequencies immediately from $t = 0$ because the modes take time to manifest in $B_{\rm rms}(t)$. Instead, we use the following time intervals: $t \in [120,300]$ for runs far from mode transition points and $t \in [820,1000]$ for runs near the transition. As a comparison, the reference studies extend simulations much more — up to $t_{\rm end}^{\rm ref} \simeq 6000$ — to achieve better mode separation. However, the selected time intervals in \textsc{Swift} are sufficient to accurately measure growth rates and oscillation frequencies.

\begin{figure}
    \centering
    \includegraphics[width=1\linewidth]{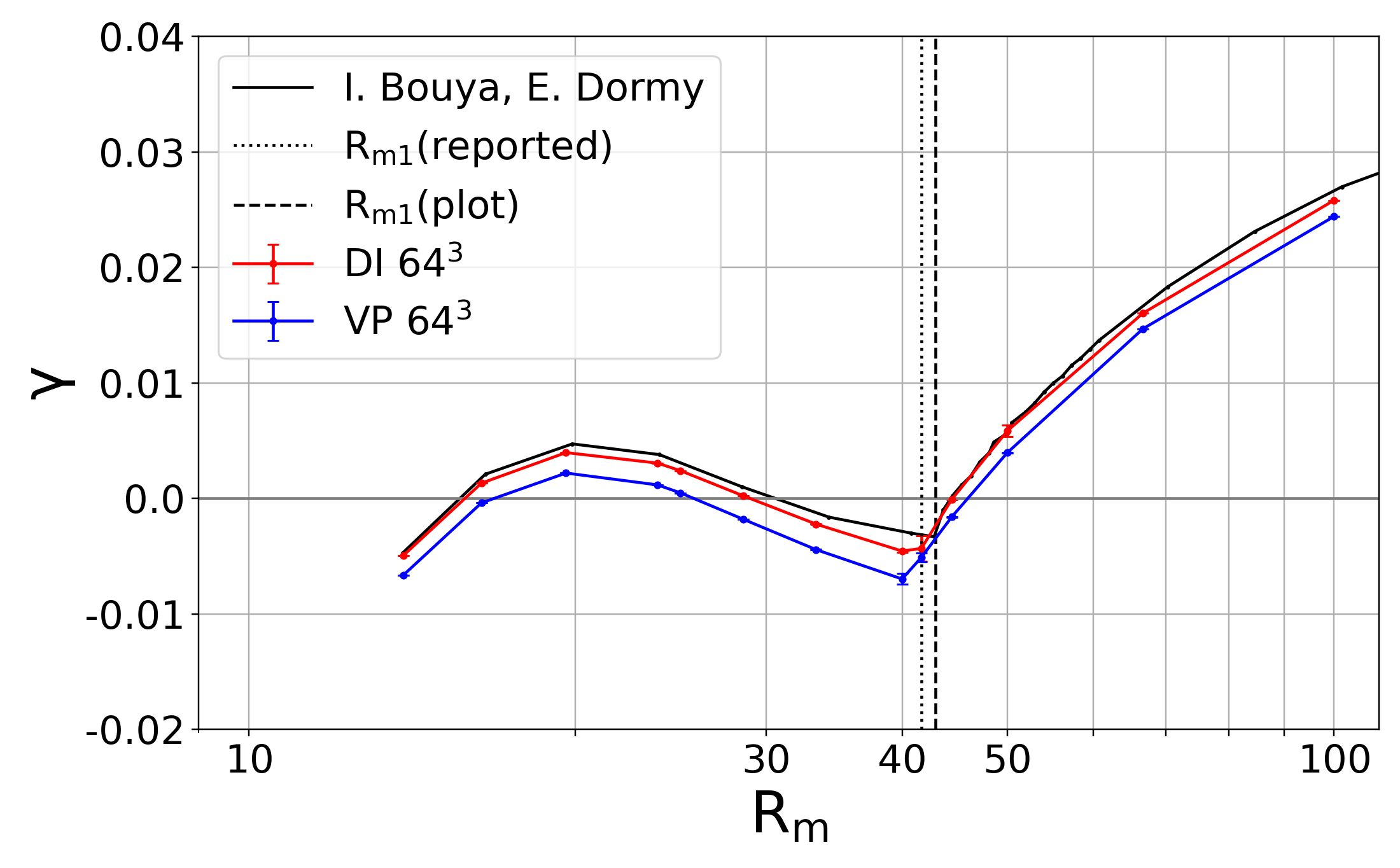}
    \includegraphics[width=1\linewidth]{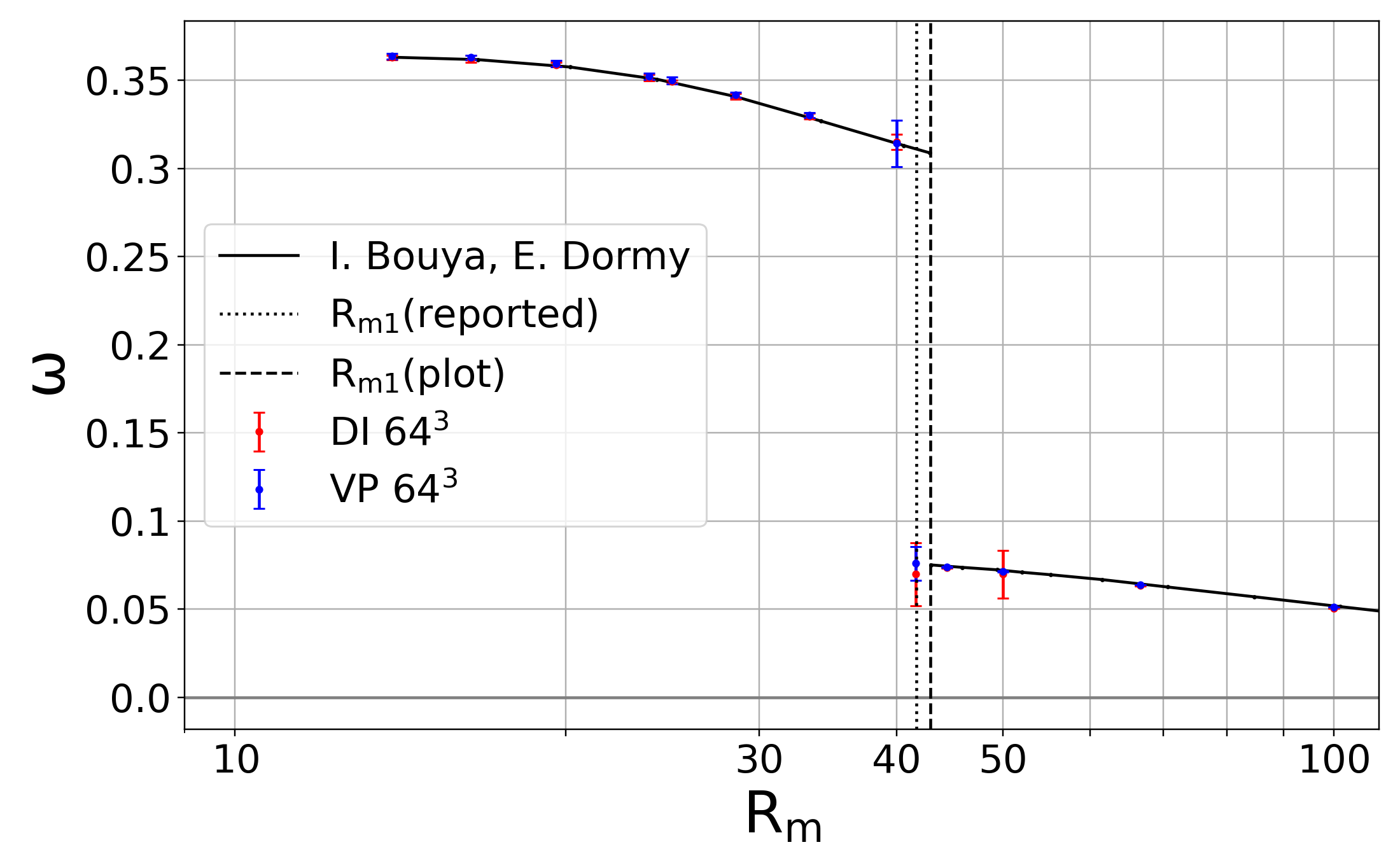}
    
    \caption{Growth rate (top plot) and frequency (bottom plot) vs magnetic
      Reynolds number for \citet{Bouya2012RevisitingTA} (converted for our
      setup) and for \textsc{Swift} MHD implementations. The \textsc{Swift} code
      reproduces oscillatiry MF growth in $R_{\rm m1} \in [10,100]$. The mode
      transition in \textsc{Swift} happens in range $R_{\rm m}\in
      [40,41.67]$. The reference transition point is $R_{\rm m1}^{\rm ref}\in
      [41.65,41.74]$, however, we find this value inconsistent with the
      transition from the plots the reference provides, $R_{\rm m1}^{\rm
        plot}\in [42.88,43.01]$, from the postion of the knee in growth rate
      graph and the step in frequency graph (Figures 1,3 from
      \citet{Bouya2012RevisitingTA} ). Error bars in growth rate and frequency
      originate the calculation methods.}
    \label{fig:ABC_growth_frequency}
\end{figure}

We conducted simulations for magnetic Reynolds numbers in the range $R_m \in [15,100]$. In all runs, we observe oscillatory growth of the magnetic field. The growth rate as a function of $R_m$ is expected to have a positive region for $R_m \in [15,30]$ and $R_m > 40$. The top plot in Figure \ref{fig:ABC_growth_frequency} shows the growth rate versus magnetic Reynolds number for $64^3$ particles in \textsc{Swift}, along with the rescaled results of \cite{Bouya2012RevisitingTA} for $v_{\rm rms} = 1$, as described earlier. Note that in the reference, the resolution was varied with $R_m$, with the minimal value being $64^3$). All MHD implementations in \textsc{Swift} follow the growth rate trends reported in the reference solutions. The critical magnetic Reynolds numbers for the onset of dynamo action are found to be $R_{\rm m}^{\rm crit,1}(\rm DI)\simeq 15.6$ and $R_{\rm m}^{\rm crit,1}(\rm VP)\simeq 16.7$, which are slightly larger than the reference value $R_{\rm m}^{\rm crit,1} \simeq 15.5$ reported in Table 1 of \cite{Brandenburg_2020}.

Similar to the Roberts Flow I results, both DI and VP implementations slightly underestimate the growth rates when compared to \cite{Bouya2012RevisitingTA}, with the VP method showing a somewhat larger deviation. Nonetheless, both models successfully reproduce the expected qualitative behavior, and the discrepancies remain small, with the growing modes appearing in the anticipated regimes.

The lower panel of Figure \ref{fig:ABC_growth_frequency} illustrates the oscillation frequency of the modes as a function of $R_{\rm m}$. The frequencies for both DI and VP implementations closely match the reference results outside the mode transition region. However, the transition occurs at $R_{\rm m1}^{\text{Swift}} \in [40,41.67] $, a value close to the reference transition at $R_{\rm m1}^{\rm rep} \in [41.65, 41.74]$. While the reported reference values overlap with \textsc{Swift}’s results, a closer inspection of their growth rate knee position and frequency evolution suggests that the actual transition happens at $R_{\rm m1}^{\rm plot} \in [42.88,43.01]$. This value differs slightly from their reported range and does not overlap with the transition values obtained for \textsc{Swift}.

As with Roberts Flow I, we expect a characteristic spatial distribution of the magnetic field. However, due to the oscillatory growth, these features evolve over time. To analyze this, we examined isosurfaces where $|\textbf{B}| / B_{\rm rms} = 3$ at the time of peak $B_{\rm rms}(t)$.

\begin{figure}
    \centering
    \includegraphics[width=\linewidth]{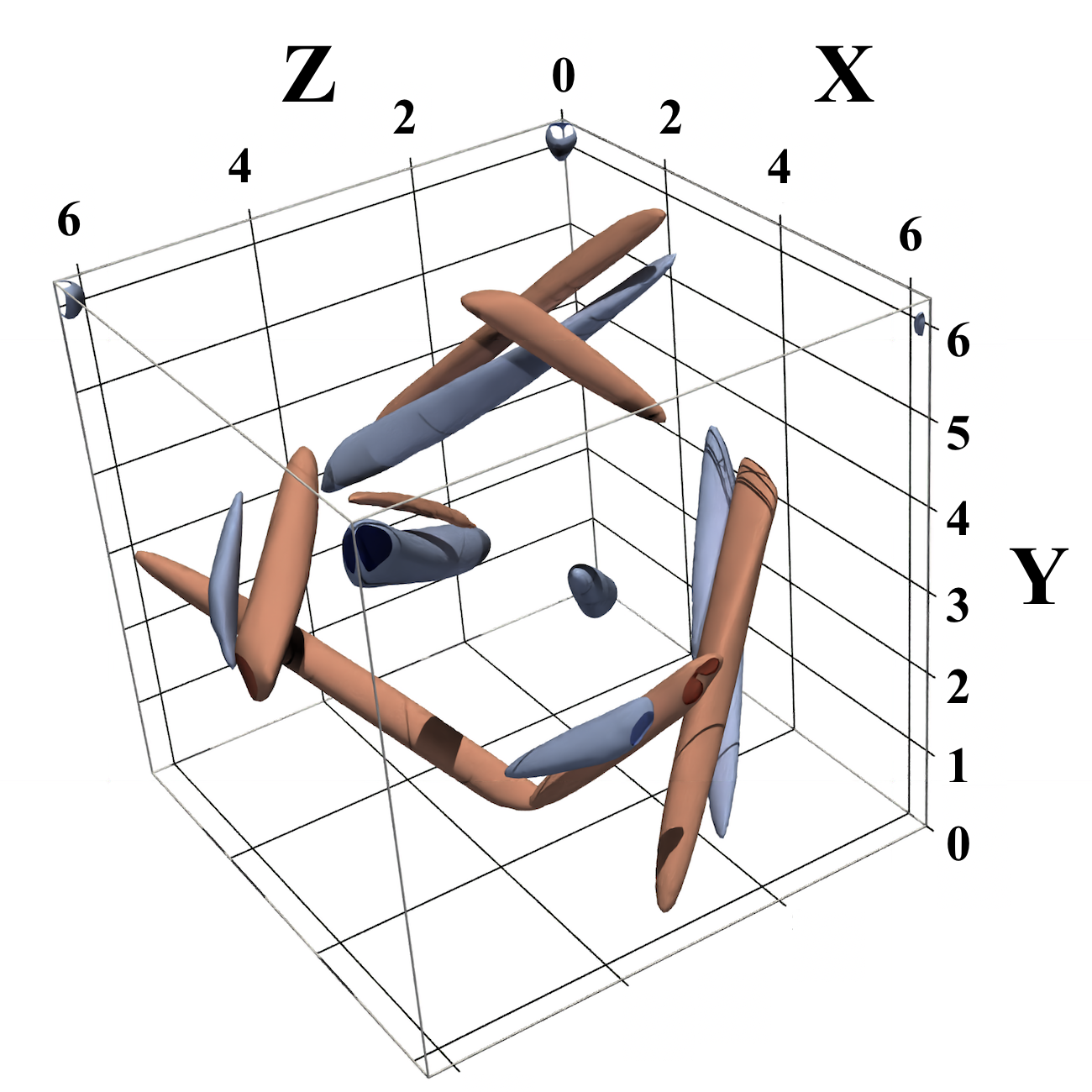} 
    \caption{Isosurfaces for $|B|/B_{\rm rms} = 3 $ magnetic field for the run
      with DI at $R_{\rm m}=100$. Color denotes sign of $B_{\rm x}$, red is
      positive. Tube-like structures are visible, similar to ones observed from
      Figure 7 from reference \citet{Bouya2012RevisitingTA}, although they run
      at $R_{\rm m}\simeq752$ ( where we multiplied $R_{\rm m}$ by $\sqrt{3}$ to
      transform to $v_{\rm rms}=1$ system). Paraview was used to visualise the
      3d isosurfaces.  }
    \label{fig:ABC_cigars}
\end{figure}

According to \cite{Bouya2012RevisitingTA}, the isosurfaces of
$\textbf{B}$ should form diagonal “magnetic field cigars” with
opposing magnetic field directions.
For the $64^3$ DI run, this cigar-like structure is indeed observed,
as shown in Figure \ref{fig:ABC_cigars} (for $R_{\rm m} = 100$).
The same structure also appears in VP runs (not shown).

\section{Resolution and overwinding problem}
\label{sec:overwinding}
Hving established that the MHD implementation in the \textsc{Swift} code can
reproduce known results on kinematic dynamos, we now turn our attention to the
link between resolution and resistivity. The discussion below follows \cite{Charbonneau2012-bx} with application to the finite resolution
scale introduced in simulations.

\subsection{Roberts Flow I minimal resistivity}

The magnetic field pattern breakdown the ideal MHD limit ($\eta \rightarrow
0,~R_{\rm m}\rightarrow\infty$) and its dependence on resolution (Figure
\ref{fig:RF1_overwinding}) suggests a connection between the simulation
resolution scale,(in our case, the smoothing length, $h$) and the
resistivity. In the Roberts Flow I test, we have 4 vortices winding up and
resistivity diffusing the magnetic field. To form a steady pattern the
resistivity should thus balance the induction term in MHD equations. 

To understand why the pattern breakdown appears, let us consider a simpler
version of the problem, where there is one single vortex with $v_z=0$ and a
magnetic field confined to the plane. The fluid movement in the vortex of size
$L_{\rm v}$ and with root mean square velocity $v_{\rm rms}$ will drag and wind
the magnetic field thus increasing the magnetic field gradients. This process is
characterised by a circulation timescale:
\begin{equation}
    t_{\rm c} \sim \frac{L_{\rm v}}{v_{\rm rms}}.
\end{equation}
If the magnetic field gradients have typical scale of $l_{\rm B}$ the diffusion
will act on characteristic times:
\begin{equation}
    t_{\rm d} \sim \frac{l_{\rm B}^2}{\eta}.
\end{equation}
A steady magnetic field pattern can form if the balance occurs, \emph{i.e.}
$t_{\rm c} \sim t_{\rm d}$. This is possible when the characteristic scale of
the gradient is:
\begin{equation}
    l_{\rm B} \sim \sqrt{\frac{\eta L_{\rm v}}{v_{\rm rms}}}.
\end{equation}
Since our SPH simulation can't resolve gradients smaller than (or of order of)
the smoothing length $h$, the timescale balance is absent for such gradient. In
this situation, the resistivity can thus, in principle, not counteract the
winding, leading to the magnetic field pattern breakdown. Alternatively, this
balance can also be thought as a magnetic field cascade that can or cannot be
countered by cut in the spectrum from the action of the physical (Ohmic)
resistivity term.

Estimates for the minimal resistivity and the maximal magnetic Reynolds number
that can reached within the simulation are thus:
\begin{equation}
    \eta_{\rm min} \sim \frac{h^2 v_{\rm rms}}{L_{\rm v}}, \hspace{0.5cm} R_{\rm m}^{\rm max} \sim \frac{v_{\rm rms} }{\eta_{\rm min} \cdot k_{\rm f}} \sim \frac{L_{\rm v}^2}{\pi h^2},
\end{equation}
where $k_{\rm f}\simeq \frac{2\pi}{2 L_{\rm v}}$ is flow wave vector was used.
This expression relates $R_{\rm m}^{\rm max}$ to how well vortices are resolved
in terms of our resolution scale $h$. We call the limiting resistivity the
\emph{overwinding resistivity}.

\begin{table}
  \centering
  \caption{Overwinding resistivity for $v_{\rm rms}=1$ and magnetic Reynolds
    number for the Roberts flow I test.
    }
  \label{tab:RF_minimal_resistivity}
  \begin{tabular}{lccr} 
    \hline
	{\rm Resolution} & h & $\eta_{\rm min}$ & $R_{\rm m}^{\rm max}$ \\
	\hline
	$16^3$ & 0.54 & $8\cdot 10^{-2}$ & 12\\ 
        $32^3$ & 0.27 & $2\cdot 10^{-2}$ & 50  \\
        $64^3$ & 0.13 & $5\cdot 10^{-3}$ & 200 \\
	\hline
  \end{tabular}
\end{table}

Applying this framework to the Roberts flow I case, we can compute the minimal
resistivity that our method can correctly evolve. The Values for the runs at
different resolutions are given in Table \ref{tab:RF_minimal_resistivity}. We
expect for the pattern to be only slightly affected when $\eta\gg\eta_{\rm min}$
and destroyed if $\eta\ll\eta_{\rm min}$. On Fig.~\ref{fig:RF1_overwinding}, we
showed the magnetic field patterns for the runs with $\eta=0.01$. At the lowest
resolution($N=16^3$, left) the pattern is highly distorted. When the resolution
is increased to $N=32^3$ and $N=64^3$, the pattern becomes more symmetric and
develops thinner features. As expected from the analysis above, at zero
resistivity (right-most panel) the balance cannot take place in principle at any
resolution. The magnetic field gradients thus reach the resolution scale and the
pattern disappears. 

Similarly, the analysis of the growth rate as a function of resistivity
(Fig.~\ref{fig:RF1_growth}) confirms that for the runs at a resolution $N=64^3$,
the large deviations of the growth rate and increase in growth rate fluctuations
happen around $\eta\sim 2-5\cdot10^{-3}$. This value is in good agreement with
the predicted overwinding resistivity value from Table
\ref{tab:RF_minimal_resistivity}.

\subsection{The overwinding trigger}
\label{ssec:overwinding_test}

For arbitrary types of flows present in astrophysical and cosmological
applications the defining a minimal resolvable resistivity or Reynolds number is
a challenging problem. We attempt to construct such a trigger here based on the
considerations exposed above.

The magnetic field gradients are governed by source terms in induction equation, which can increase them:
\begin{equation}
    \textbf{S}_{\rm ind} = \Omega_{\rm str}+\Omega_{\rm Dedner}
\end{equation}
Diffusion sources decrease the gradients:
\begin{equation}
  \textbf{S}_{\rm diff} =\Omega_{\rm Ohm}+\Omega_{\rm AR}
\end{equation}

The gradients in the SPH simulations are bound due to the
resolution scale. This rough estimate of the maximal gradient that can be
resolved typically holds:
\begin{equation}
    |\Delta \textbf{B}|\leq \frac{2|\textbf{B}|}{h^2}.
    \label{laplacian_bound}
\end{equation}

Therefore, there exists a limit on the diffusive source term in the induction equation, which in turn sets an upper bound on the achievable magnetic Reynolds number when magnetic field gradients reach resolution scale. 
If the induction source term,
$\textbf{S}_{\rm ind}$, counteracts the diffusion (\emph{i.e.} $\textbf{S}_{\rm
  diff}\cdot \textbf{S}_{\rm ind} <0$) and $\textbf{S}_{\rm ind} > \textbf{S}_{\rm ind}^{\rm max} \simeq \frac{(\eta+\eta_{AR})}{\rho} \cdot \frac{2|\textbf{B}|}{h^2}$, then in the presence of magnetic field cascade the magnetic fields will inevitably reach resolution scales, or $|\Delta \textbf{B}| \simeq \frac{2 |\textbf{B}|}{h^2}$.
 This will result in overwinding issues similar to what was found in
the Roberts Flow I.

To monitor this issue we define an \emph{overwinding trigger} as follows:
\begin{equation}
    OW = \frac{|\textbf{S}_{\rm ind}|}{|\textbf{S}_{\rm diff}|}\cdot\frac{1}{2}\big(1-\cos(\textbf{S}_{\rm ind},\textbf{S}_{\rm diff})\big)\cdot \frac{h^2 |\Delta \textbf{B}|}{2|\textbf{B}|},
    \label{OW_definition}
\end{equation}
where $\textbf{S}_{\rm diff}$ includes all diffusive sources in the simulation
(physical and artificial)\footnote{Note that for this study we choose to include
the Dedner cleaning term into the induction source in the trigger
(eq.~\ref{OW_definition}).}. We expect the trigger value to be large, $OW\gtrsim
10$, if there is significant overwinding and $OW\lesssim 10^{-1}$ if the
resistivity manages to counteract the winding locally and thus separate the
cascade from the resolution scale.

The observed behaviour for the magnetic fields on Fig.~\ref{fig:RF1_overwinding}
can now be reinterpreted in terms of this $OW$ metric. When the resolution is
increased, the $OW$ trigger will decrease too since the magnetic field gradients
can reach a smaller scales and thus result in a much larger $\eta \Delta
\textbf{B}$. In the highest resolution case ($N=64^3$), this is enough to separate
the magnetic field from the resolution limit. And, as expected, for the case of
zero resistivity nothing prevents the magnetic fields from reaching the
resolution scales resulting in the vortex pattern destruction.

\begin{figure}
    \centering
    \includegraphics[width=1\linewidth]{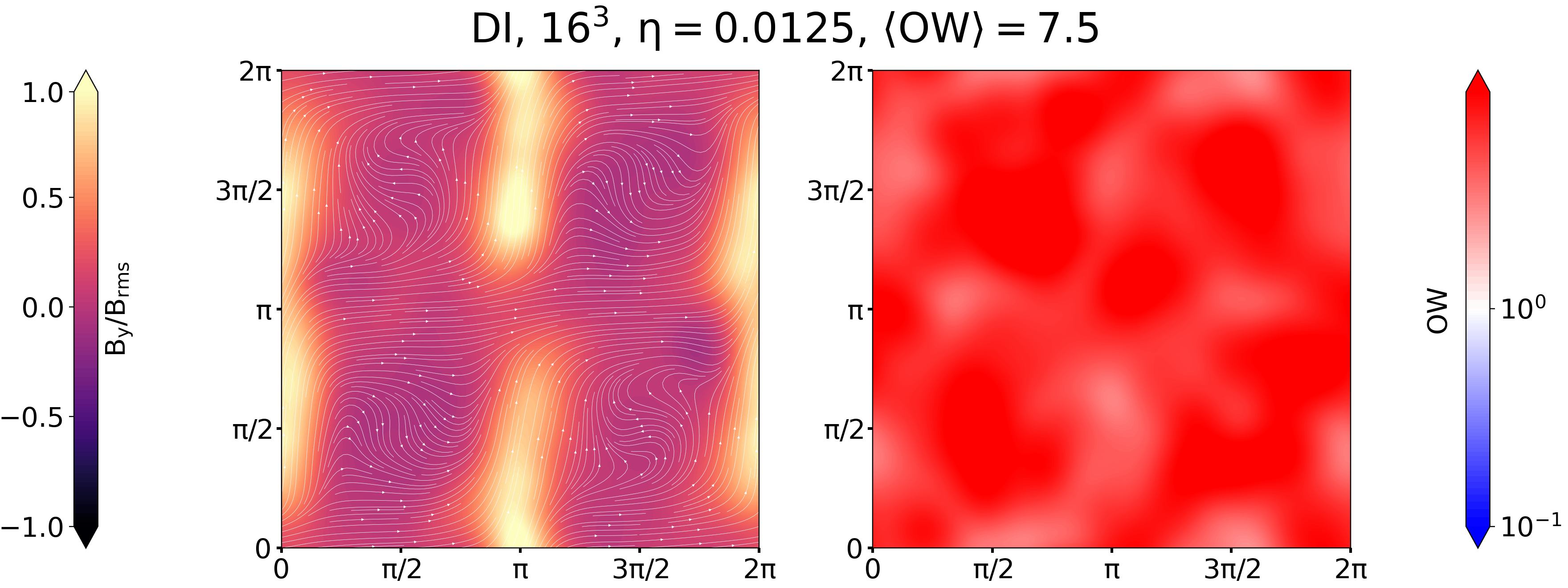}
    \includegraphics[width=1\linewidth]{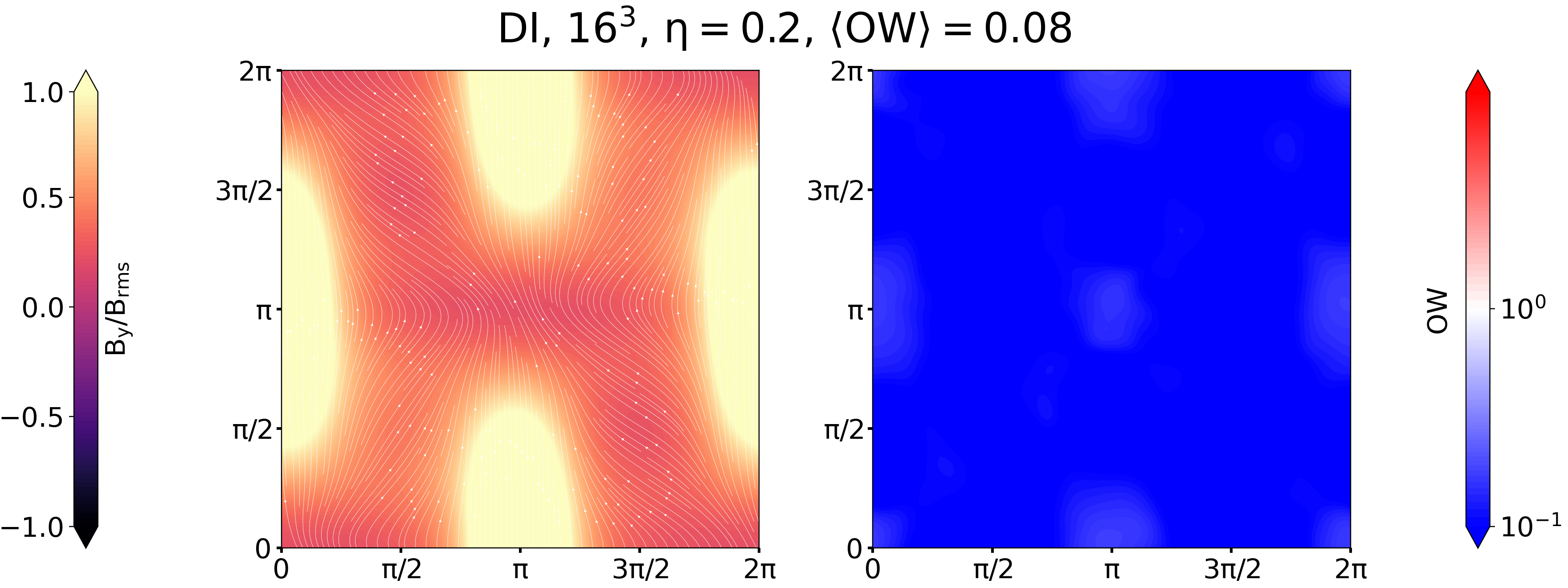}
    \includegraphics[width=1\linewidth]{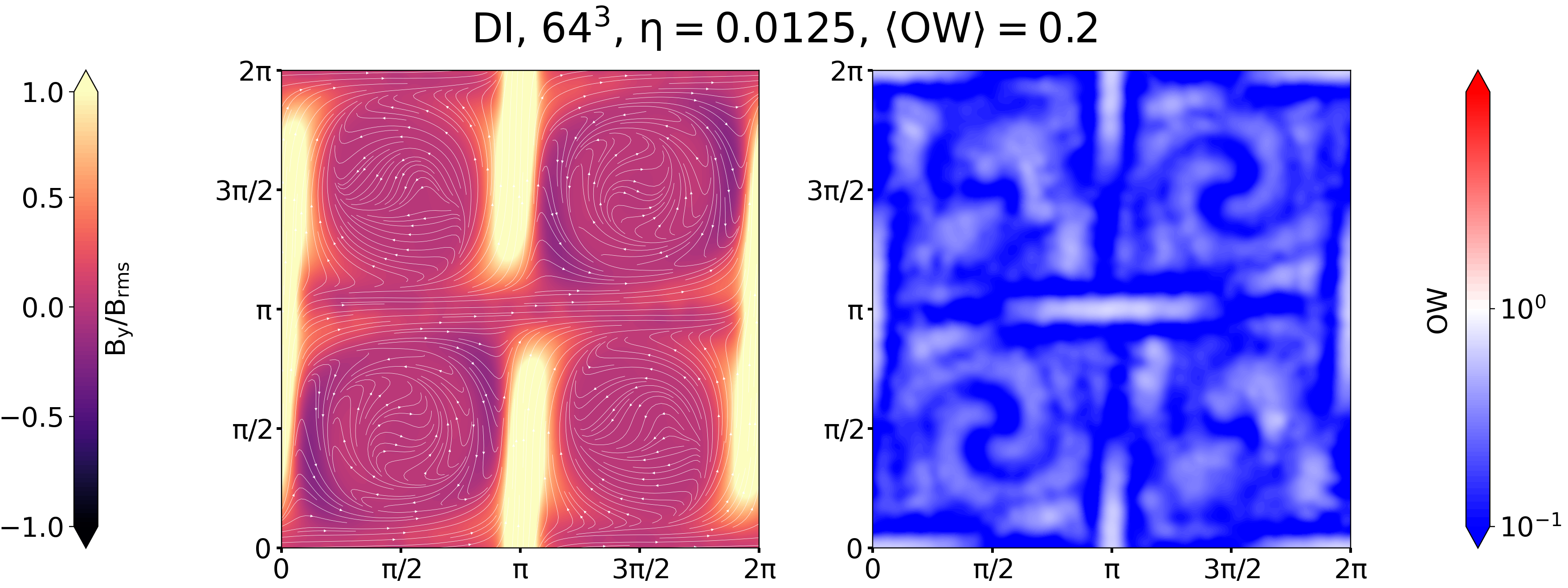}

    \caption{$B_{y}/B_{\rm rms}$ with magnetic field streamlines (left) and
      Overwinding trigger maps (right) for the Roberts flow I test in $xy$ plane at $t=75$, with heights adjusted to show the same feature. The top set of plots show the unresolved regime with
      large $OW$ trigger values ($\langle OW\rangle \gtrsim 10$). There are two
      ways to enter the properly resolved regime (where $\langle OW\rangle
      \lesssim 10^{-1}$): with resistivity increase (middle set of plot) or with
      resolution increase (bottom plot). Note also that the $OW$ metric tends to
      have higher values inside shear zones and inside vortices}
    \label{fig:OW_trigger_example}
\end{figure}


We conclude this section with a visual example of the behaviour of the
overwinding trigger. If trigger the values are large, $OW\gtrsim10$, the
resistivity plays only a small role in the overall evolution of the field. For
this regime, $OW$ follows the scaling:
\begin{equation}
    OW\sim h^2\eta^{-1}, \hspace{0.5cm} OW\big|_{\rm h=const}\sim N_{\rm
      p}^{-2/3} \eta^{-1},
    \label{OW_scaling}
\end{equation}
where the second equation holds for a fixed resolution case (i.e. where there is
a direct link between $N$ and $h$). This scaling can be used to estimate the
required resolution or resistivity needed in a simulation to prevent overwinding ($OW\lesssim1$).

Such a use case is depicted in Fig.~\ref{fig:OW_trigger_example} for the Roberts
flow test. The left panels show the $B_y/B_{rms}$ magnetic field configuration in the $xy$ plane at time $t=75$ 
with slice heights adjusted to see the same feature whilst the right panels show the
corresponding value of the $OW$. The top row corresponds to a setup where the
field configuration cannot be properly resolved with $16^3$ particles and this
value of the resistivity. Either a resistivity increase from
$\eta=0.0125$ to $\eta=0.2$  (1st to 2nd panel) or a resolution increase form $16^3$ to $64^3$
particles (1st to 3rd panel) are sufficient to decrease the trigger value to the
regime where the vortices are fully resolved by the simulation. Of course,
changing the value of the physical resistivity is not always a practical option
as its value can be set by physical considerations. Increasing the resolution is
then the sole option to obtained a well-behaved field with resolved kinematic
dynamo growth.

The computational cost incurred then increasing resolution can be estimated as
$\sim N \cdot T_{\rm sim}/\Delta t$, with $T_{\rm sim}$ the final time
of the simulation and $\Delta t$ the time-step size. This latter quantity is
itself $\Delta t \sim \frac{h}{c_{\rm h}}$, where $c_{\rm h}$ is signal
propagation speed (sound speed typically or Alfven wave speed for MHD
scenarios).  Assuming a constant propagation speed, the full cost is then $\sim
N_{\rm p}^{\frac{4}{3}}\sim (OW\cdot \eta)^{-2}$. Thus, targeting a decrease of
the $OW$ value by an order of magnitude will cost two orders of magnitude more
computational costs. On the other hand, an order of magnitude increase in the
resistivity constant value will lead to the same magnitude change in $OW$.

Therefore, the most effective way to prevent overwinding and avoid excessive damping of relevant physical features at a given resolution is to adjust the resistivity such that $OW \simeq 1$.

\section{Cosmological MHD simulations}
\label{sec:CosmoVolume}

\begin{figure*}
  \centering
  \includegraphics[width=1\linewidth]{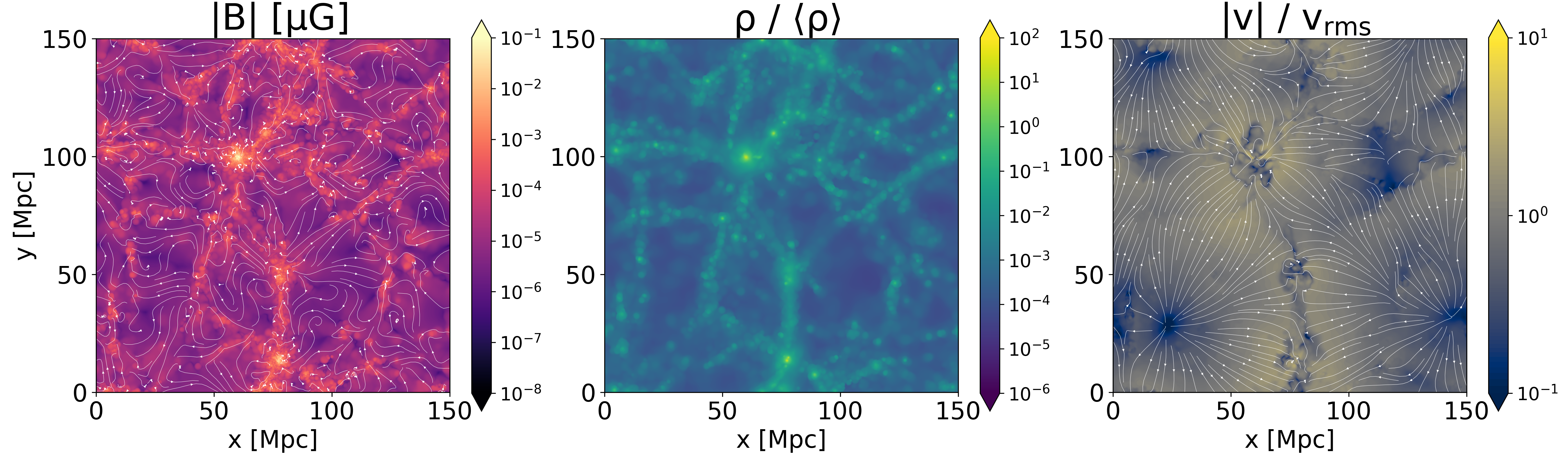}
  \vspace{-0.3cm}
  \caption{Magnetic field in micro-Gauss with field streamlines (left), gas
    over-density (center) and ratio of velocity to root-mean-square velocity of
    the gas with flow streamlines (right) for adiabatic cosmological run with
    zero resistivity at redshift $z=0$ with $128^3$ particles using the direct
    induction SPMHD implementation in the \textsc{Swift} code. 
      The density map shows, as expected, the formation of voids and
    filaments, while the velocity profile shows the flow of gas from voids into
    filaments and high density nodes. The magnetic field amplitude distribution
    mostly follows the density.} \label{fig:Cosmo_DI_IMHD_B_rho_v}
\end{figure*}

Having demonstrated, that the MHD implementation in \textsc{Swift} code
reproduced the features of the Roberts flow and ABC flow kinematic dynamo tests
within the resolution-dependent Reynolds number window, we now want to
demonstrate its capability to capture the basic amplification processes
associated with dynamo action. In this section we explore simulations solving
cosmological MHD equations, with a focus on the dynamics without considering
sub-grid physics models (i.e. so-called \emph{adiabatic} simulations).

Cosmological MHD simulations involve various flow types during structure
formation, which can lead to magnetic field amplification through gravitational
collapse and the stretching of magnetic field lines. Cosmological runs without
resistivity can suffer from the ``overwinding'' problem discussed in the last
section, where the magnetic field becomes excessively stretched. To mitigate
this, we perform additional simulations with non-zero constant physical
resistivities: first, with a typical value found in galaxy clusters, $\eta \sim
6 \times 10^{27} \, {\rm cm}^2/{\rm s}$, and then with higher resistivity values
to further minimize overwinding.

\subsection{Simulation setup}

Cosmological simulations with a side-length of $150 \, {\rm Mpc}$ were conducted
using MHD and adiabatic gas evolution, that is without including feedback or
cooling physics. We perform simulations with $2\times 64^3$ and $2\times128^3$
particles. This leads to a gas particles mass of $m_{\rm gas} = 8\cdot 10^{10}
{\rm M}_{\odot}$ and $1\cdot 10^{10} {\rm M}_{\odot}$ respectively.  The initial
conditions for baryons and dark matter were generated using \textsc{Monofonic}
code \citep{Hahn_2020} at a starting redshift of $63$. We adopt the same
cosmology as the \textsc{FLAMINGO} project \citep{Schaye_2023}\footnote{
$\Omega_{\rm CDM}=0.2574$, $\Omega_{\rm b} = 0.0486$, $\Omega_{\Lambda} =
0.693922$, $h = 0.681$, $n_s = 0.967$, $A_s = 2.099 \cdot 10^{-9}$, }.  The initial
magnetic field was generated as a Beltrami field (eq.~\ref{Beltrami_field}),
with 10 waves along one axis of the box, and a root-mean-square code-comoving
magnetic field strength of $\rm B_{\rm rms}^{comov}(z_0) = 10^{-6} \mu G$, which
corresponds to $\rm B_{\rm rms}(z_0) = 2.97\cdot 10^{-2} \mu G$ in physical magnetic
field. This field is uncorrelated with the density structure in the ICs and does
not represent a physical scenario. Note that the Beltrami field configuration
was chosen to make initial conditions reproducible for runs using a vector
potential scheme, which we will explore in future studies.

\begin{figure*}
    \centering
    \includegraphics[width=1\linewidth]{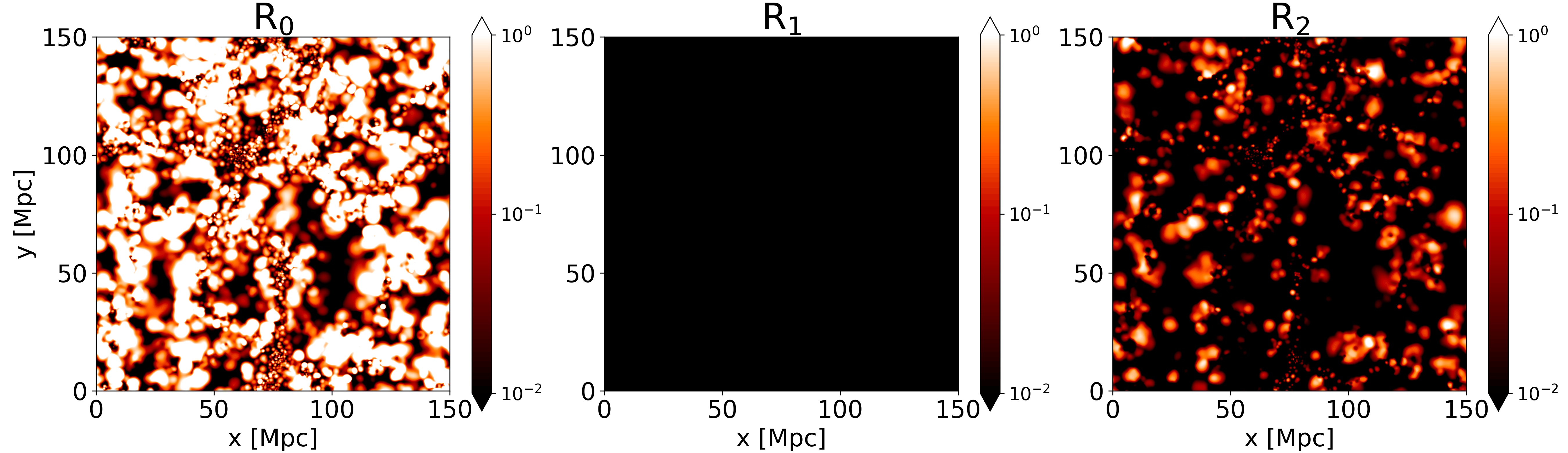}
    \vspace{-0.3cm}
    \caption{Divergence error metrics (eqs. \ref{error_R0} -- \ref{error_R2}) for
    our adiabatic cosmological run with zero resistivity at redshift $z=0$ and
    $128^3$ particles using the direct induction SPMHD implementation
    in \textsc{Swift}. The $R_0$ metric indicates that there are regions with
    large divergence errors, however, the $R_2$ error metric lights up less
    showing that at some places with large $R_0$ the physical field is more
    significant and thus the divergence error may not influence the dynamics of
    the field. Therefore, there are large magnetic field gradients at resolution
    scale. Since the $R_1$ is below our noise cut everywhere, we expect no
    unphysical (monopole) force acting on the matter evolution. The $R_0$ errors
    have large volume filling fraction and concentrate at density
    gradients.}
    \label{fig:Cosmo_DI_IMHD_errors}
\end{figure*}

\subsection{Ideal MHD cosmological runs}

We start out analysis by an overview of the general properties of the gas
distribution at $z=0$ for our higher-resolution run.

In Fig.~\ref{fig:Cosmo_DI_IMHD_B_rho_v}, we show the spatial distribution of an
infinitely thin slice in the xy-plane, depicting the magnetic flux density (in
${\rm \mu G}$, left panel), the matter over-density (middle panel), and the
ratio of local velocity to the root-mean-square velocity across the simulation
box (right panel). Streamlines in the magnetic flux density plot illustrate the
geometry of the magnetic field in the xy-plane, while streamlines in the
velocity plot represent the gas flow. The magnitude of the magnetic field
strength largely follows the gas density, with denser regions exhibiting
stronger magnetic fields. Gas density and velocity profiles follow the usual
pattern expected from such cosmological simulations: gas forms voids and
filaments, with velocity profile showing the gas flux from low density regions
into dense filaments. The low-density areas display a smoother magnetic
field. In the filaments, the magnetic field strength is non-uniform, both in
amplitude and direction. Additionally, the velocity streamlines reveal
stagnation-point-like flows, similar to those observed between vortices in the
Roberts flow I. These flows, found in and along the filaments, suggest that
magnetic field amplification may occur not only due to gravitational collapse
but also as a result of stretching within these flows.

We now turn out attention to the value of the three error metrics we introduced
in Sec.~\ref{ssec:error_metrics}. On Fig.~\ref{fig:Cosmo_DI_IMHD_errors} we show
the spatial distribution, in the same plane as for the previous figure, of
$R_0$, $R_1$, and $R_2$ (eqs. \ref{error_R0} -- \ref{error_R2}). The SPH noise
cut discussed in Sec.~\ref{ssec:error_metrics} has been applied using the noise
estimates from equation \ref{errors_noise_cut}. We show the error in the range
$R_i \in [10^{-2}, 10^{0}]$.

As can be seen, a significant fraction of the simulation volume exhibits large
divergence errors ($R_0$). However, when compared to the density and magnetic
field distributions in Fig.~\ref{fig:Cosmo_DI_IMHD_B_rho_v}, both the centers of
low-density regions and the areas within filaments show lower error levels than
the regions at the borders between low and high-density areas, where large
density gradients are found. This indicates that the error is possibly driven by
poorer quality gradient operators in such regions. The other divergence error
metrics show much lower of error. The $R_1$ metric is below the noise cut across
the entire simulation, indicating that there are no significant magnetic
monopole forces acting on the particles in this setup. The $R_2$ metric, which
estimates the monopole component of the magnetic field relative to the physical
current, shows significantly fewer errors than $R_0$ overall, typically at a
level smaller than $R_2 < 10^{-1}$. This, by design, also suggests the presence
of large gradients in both the physical and monopole components of the magnetic
fields at the resolution scale, with only some regions (indicated by $R_2$)
exhibiting a significant monopole component.
 
\subsection{Non-ideal MHD cosmological runs}

\begin{figure*}
    \centering
     \vspace{0.1cm}
     \includegraphics[width=1\linewidth]{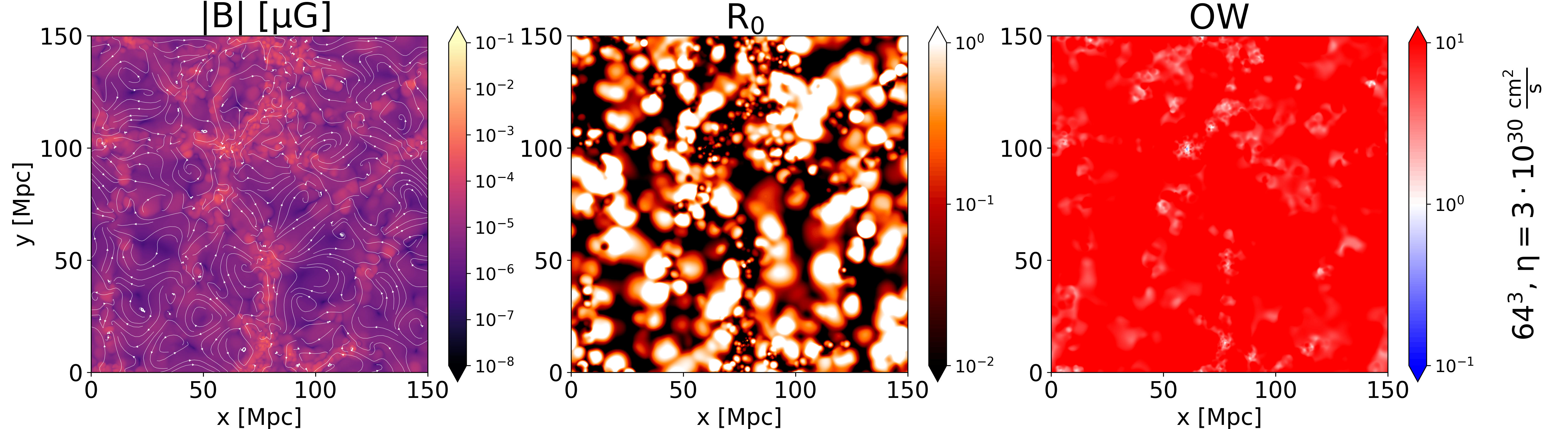}
    \vspace{0.1cm}
    \includegraphics[width=1\linewidth]{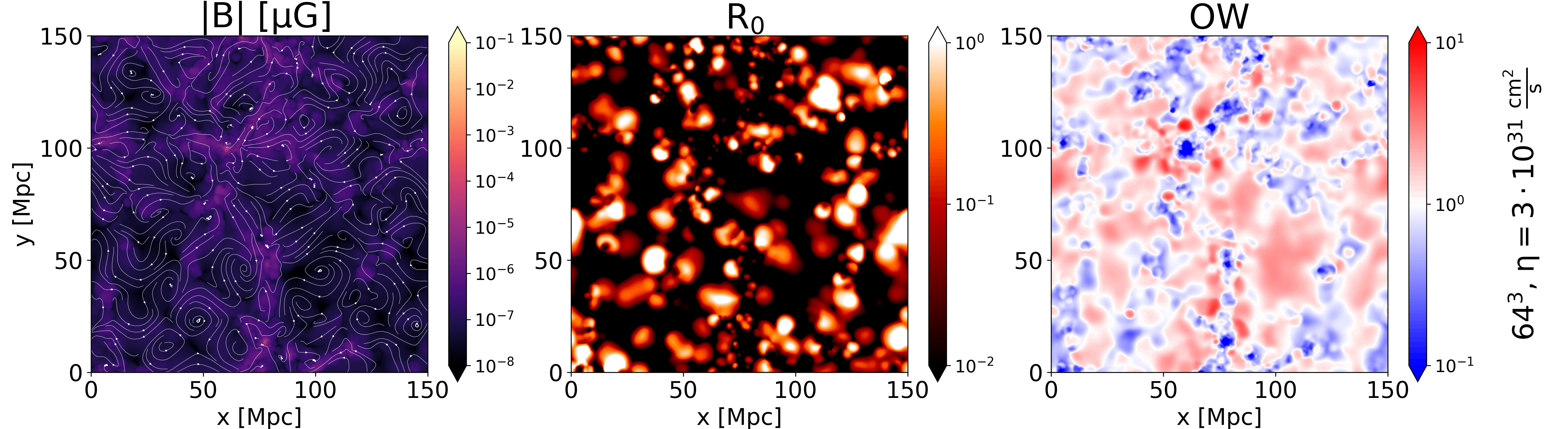}
     \vspace{0.1cm}
     \includegraphics[width=1\linewidth]{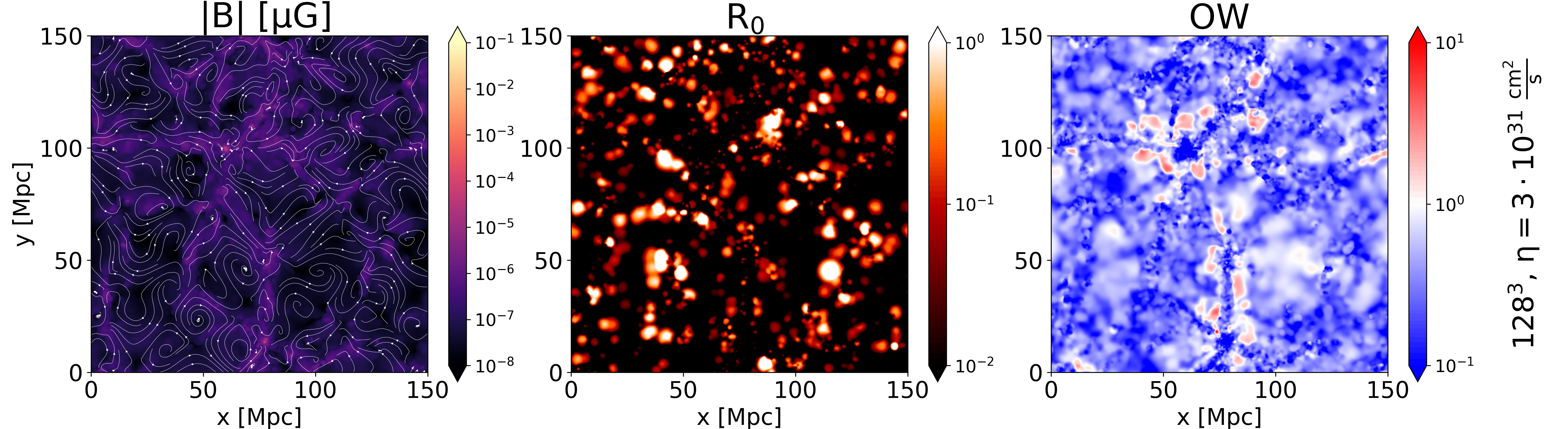}
     \vspace{0.1cm}
     \includegraphics[width=1\linewidth]{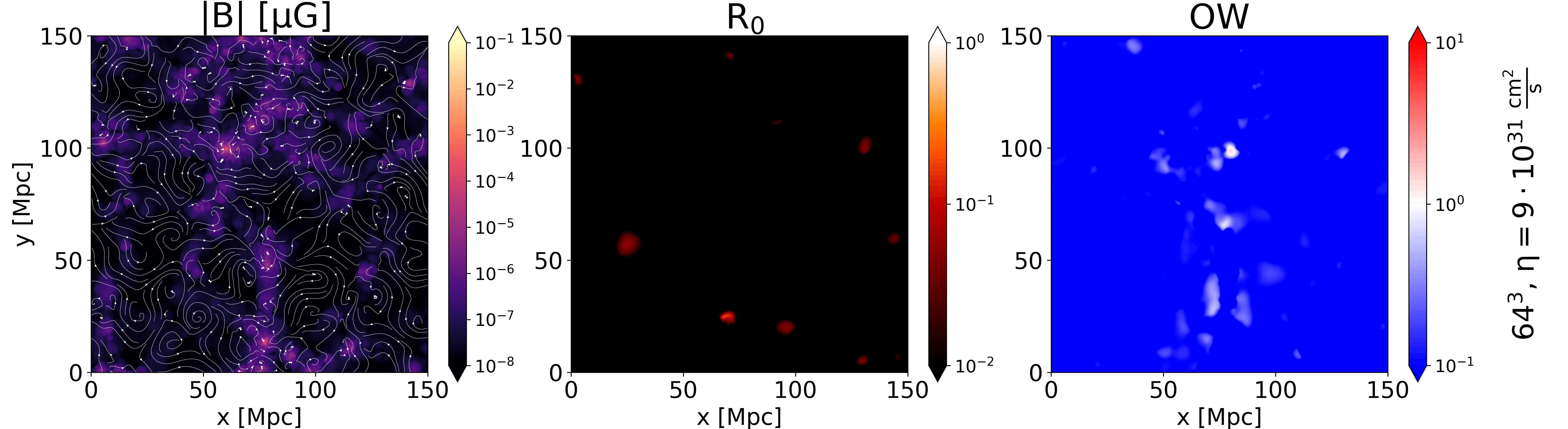}
    \vspace{-0.3cm}
    \caption{Maps of the magnetic field $B$ (in $\rm{ \mu G}$), error metric
      $R_0$, and $OW$ trigger values in our adiabatic cosmological simulations
      at $z=0$ run with the direct induction SPMHD implementation in
      \textsc{Swift} for a $64^3$ particles run with $\eta = 3\cdot
      10^{30}~\rm{cm}^2{\rm s}^{-1}$ (top row), $\eta = 3\cdot
      10^{31}~\rm{cm}^2{\rm s}^{-1}$ (second row), a run with the same resistivity but
      with $128^3$ particles (third row), and a run with $64^3$ particles but
      $\eta = 9\cdot 10^{31}~\rm{cm}^2{\rm s}^{-1}$ (bottom row).  Increasing the
      resistivity (1st to 2ns row) leads to a decrease in the $OW$ trigger
      values and a reduction in the volume-filling fraction of divergence
      errors, while also significantly decreasing the magnetic field
      magnitude. Increasing the resolution from the second to the third row
      further reduces the $OW$ trigger, though the overall $R_0$ filling
      fraction remains similar. From the second to the bottom row, a further
      increase in resistivity significantly reduces the $OW$ trigger while
      bringing divergence error levels to an acceptable range ($R_0 <
      10^{-1}$). At the same time, the magnetic field structure changes
      slightly: some filament regions experience damping, while high-density
      regions exhibit increased magnetic field strength.}
    \label{fig:Cosmo_DI_OW}
\end{figure*}


We now consider the case of cosmological simulations with Ohmic resistivity. We
run the same setup as explored above but start at $8\times$ lower mass
resolution and using a a constant physical resistivity $\eta = 6 \times
10^{27}~\rm{cm}^2{\rm s}^{-1}$; a value typical for galaxy clusters  \citep{Bonafede_2011}. For such a simulation, we found that the the $OW$
(eq.~\ref{OW_definition}) is very high everywhere. Its mean value was found to
be $\langle OW \rangle \sim 10^4$, indicating that magnetic field gradients are
under-resolved.


Since we expect the overwinding metric to follow the scaling from
eq.~\ref{OW_scaling}, two approaches can mitigate this issue and achieve a
better resolved kinematic dynamo for the fields (i.e. $\langle OW \rangle
\lesssim 1$): (1) increasing the resistivity to $\eta \gtrsim 6 \times
10^{31}~\rm{cm}^2\rm{s}^{-1}$ or (2) increasing the number of particles.


However, it is worth noting that, in a typical cosmological setup, increasing
the number of particles may introduce new substructures, which could again be
under-resolved, but on smaller scales. In regions where a mostly uniform density
is expected, such as voids, reducing the trigger values from $10^4$ requires
increasing the particle count sufficiently to decrease $h/L_{\rm B}$ by at
least a factor of 100, a challenging demand.

We showed the effect of changing resolution or the value of resistivity on the
Roberts flow I test in Sec.~\ref{ssec:overwinding_test}. We now perform the same
type of experiments in our cosmological setup, in order to demonstrate the
importance of monitoring $OW$ also in SPMHD applications beyond tests.

To this end, we conducted three more simulations at the same resolution ($64^3$
particles) but with three different resistivity values: $\eta = 3 \times
10^{30}~\rm{cm}^2\rm{s}^{-1}$, $\eta = 3 \times 10^{31}~\rm{cm}^2\rm{s}^{-1}$,
and $\eta = 9 \times 10^{31}~\rm{cm}^2\rm{s}^{-1}$, corresponding to,
respectively, the first, second, and last rows of Fig.~\ref{fig:Cosmo_DI_OW}. The
figure presents the spatial distribution of an infinitely thin slice in the
xy-plane at redshift zero, showing the magnetic flux density (in $\rm{\mu G}$,
left), the divergence error $R_0$ (including our noise cut, middle column), and
the overwinding trigger values (right column). Streamlines in the magnetic flux
density plot illustrate the field geometry. From top to bottom, both resolution
and resistivity vary across the panels, as indicated on the right of the figure.

At the lowest resistivity, $\eta \simeq 3 \times 10^{30} \ \rm{cm^2/s}$ (top
row), the $R_0$ profile reveals significant volume-filling errors, while the
overwinding trigger remains high at $\langle OW \rangle \simeq 6.95$.

With a tenfold increase in resistivity (second row of
Fig.~\ref{fig:Cosmo_DI_OW}), the overwinding trigger decreases to $\langle OW
\rangle \simeq 0.56$, which is close to the expected value of 0.69 from the OW
resistivity scaling (eq. \ref{OW_scaling}). The OW trigger maps are non-uniform,
showing higher values in low-density regions compared to high-density
regions. The volume-filling fraction of $R_0$ errors is also reduced, though
some regions still exhibit $R_0 > 1$. Additionally, a significant decrease in
magnetic field magnitude is observed, while its overall morphology remains
largely preserved.

With a further three-fold increase in resistivity (bottom row of
Fig.~\ref{fig:Cosmo_DI_OW}), the overwinding trigger $\langle OW \rangle$
deviates from the expected scaling, decreasing more rapidly to $\langle OW
\rangle \simeq 0.01$. However, regions with $OW \simeq 1$ persist near the
filament edges. The global error levels are significantly reduced, remaining
within an acceptable range of $R_0 < 10^{-1}$ throughout. Comparing the second
and bottom rows of Fig.~\ref{fig:Cosmo_DI_OW}, the magnetic field amplitude
shows different trends: it is dampened in some regions along the filaments,
while inside the filaments and at the nodes, it becomes amplified.

Finally, we increased the resolution of the simulation with $\eta = 3
~\rm{cm}^2\rm{s}^{-1}$ from $64^3$ to $128^3$ particles in order to locally
reduce the smoothing length and mitigate overwinding. The results from this
higher-resolution simulation are shown on the third row of
Fig.~\ref{fig:Cosmo_DI_OW}. The resolution change led to a global reduction of
the overwinding trigger in both high- and low-density regions, as expected from
the scaling of the metric. However, some areas near the filament edges still
exhibit large overwinding. Meanwhile, the volume-filling fraction of $R_0$
remains largely unchanged with only a small reduction. Recall though that we
found to the large divergence error to not be dynamically relevant in our
analysis of the ideal-MHD case.

\subsection{Discussion}

\begin{figure}
    \centering   
    \includegraphics[width=\linewidth]{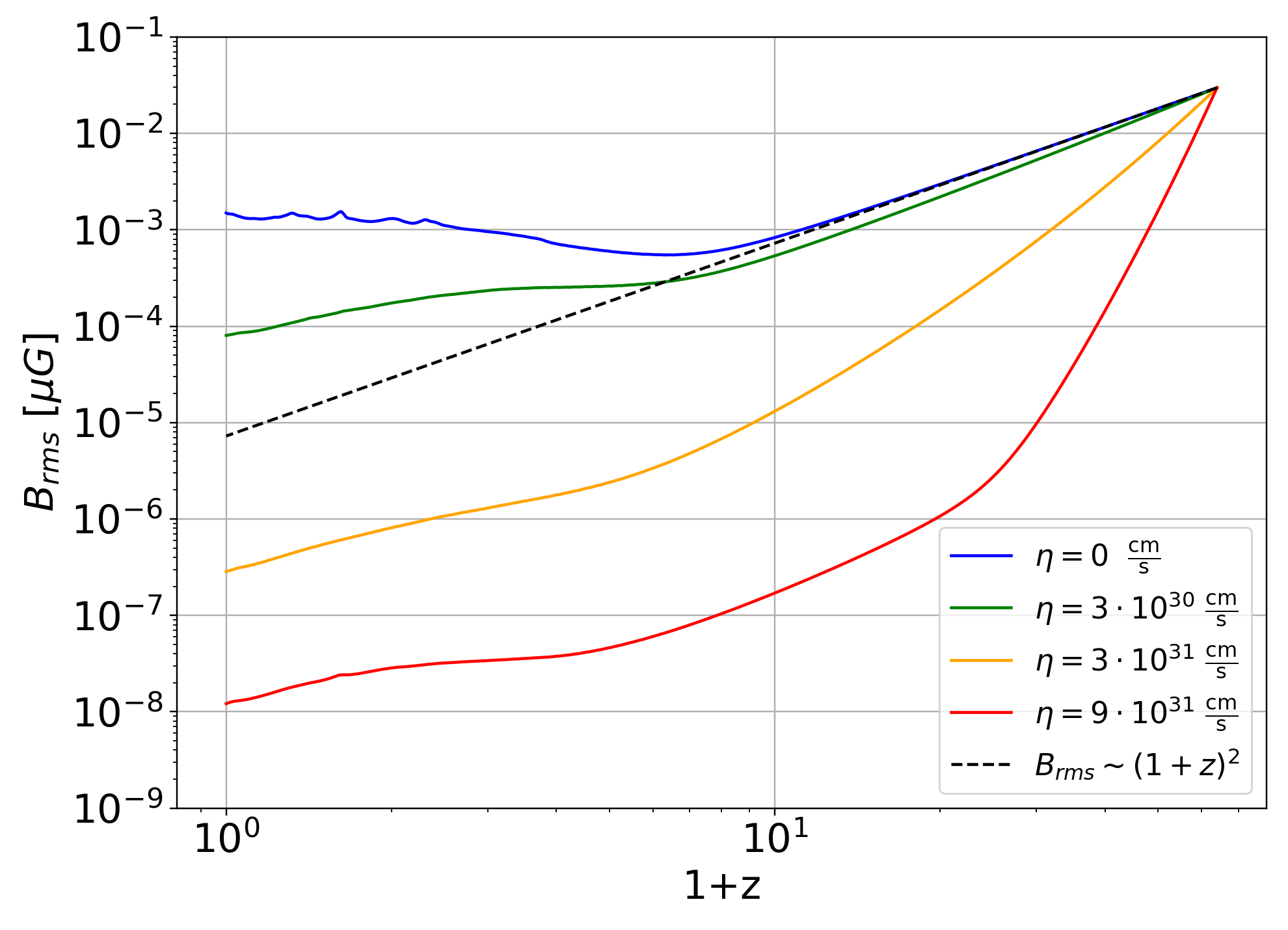}
    \vspace{-0.5cm}
    \caption{Root mean square magnetic field vs scale-factors for the adiabatic
      cosmological runs with $64^3 $ particles with $\eta = 0$; $ 3\cdot
      10^{30}$; $3\cdot 10^{31}$; $9\cdot 10^{31}$; $
      10^{33}\rm{cm}^2\rm{s}^{-1}$ (corresponding to the $z=0$ slices shown in
      Fig.~\ref{fig:Cosmo_DI_IMHD_errors} and \ref{fig:Cosmo_DI_OW}). 
      A large
      constant physical resistivity strongly dampens the initial magnetic field
      amplitude. The evolution is consistent with cosmological expansion around $z\simeq 10$, followed by amplification near $z\simeq 0$.
    }
    \label{fig:Cosmo_DI_OW_Brms_vs_z}
\end{figure}

As we just demonstrated, our \textsc{Swift}-based SPMHD adiabatic cosmological
simulations (i.e. without additional sub-grid physics) produce reasonable
magnetic field, density distributions, and velocity slices, thus demonstrating
the general reliability of the implemented MHD model. However, the accuracy of
the results is affected by large errors, which are primarily attributed to
unresolved gradients in both the physical and monopole components of the
magnetic field. These errors tend to concentrate around density gradients,
suggesting a need for additional numerical techniques to mitigate their impact.

While the addition of constant physical resistivity helps reduce both $R_0$ and
$OW$, it also significantly dampens the magnetic fields. This is, thus, an
undesirable solution in many cases.

To illustrate this, we show on Fig.~\ref{fig:Cosmo_DI_OW_Brms_vs_z} the
evolution of the root-mean-square magnetic field strength, $B_{\rm rms}$ (in
$\rm{\mu G}$), as a function of redshift for our simulations with $64^3$
particles with zero physical resistivity (blue lines), as well as for the
simulations corresponding to the first (green), second (yellow), and last (red)
rows of Fig.~\ref{fig:Cosmo_DI_OW}. The dashed line indicates the cosmological
dilution of the initial magnetic field if no MHD forces were present.

The expected solution is that of cosmological expansion line (black dashed) for $z>10$, followed by amplification from structure formation.\\

A significant magnetic field damping is observed at high redshifts ($z > 10$)
for non-zero resistivity. This occurs because, at earlier times, when the
Universe is smaller and magnetic field gradients are on smaller scales, the
constant resistivity term has a stronger effect. As a result, a substantial
initial damping occurs.

  
  Nevertheless, after the initial damping at $ z \lesssim 10 $, all runs with resistivity exhibit a similar evolution. Initially, they follow the same tilt as the evolution driven by cosmological expansion, followed then by amplification near z = 0.

A potential solution is to implement an adaptive artificial resistivity that dynamically responds to the overwinding metric, maintaining it around $OW \simeq 1$. In the case for runs above this approach could reduce excessive damping at high redshifts while still effectively controlling overwinding at z = 0. However, any such implementation should be designed to not violate astrophyiscs considerations. 
Future work should explore this adaptive method to balance error correction with physical accuracy in evolving magnetic fields.

\section{Conclusions}
\label{sec:conclusions}


In this paper, we presented the results of new \textsc{Swift} SPH MHD implementations using direct induction (DI) and vector potential (VP) methods for kinematic dynamo tests. Both implementations successfully reproduce the expected qualitative and quantitative features when compared to other codes.

For Roberts Flow I, growth rates (Figure \ref{fig:RF1_growth}) and the spatial distribution of the magnetic field (Figure \ref{fig:RF1_field_configuration}) closely match \textsc{Pencil} code results, demonstrating numerical convergence (Figure \ref{fig:RF1_convergence_log}). Additionally, we examined ABC flow, where magnetic field structures (Figure \ref{fig:ABC_cigars}), growth rates, and oscillation frequencies (Figure \ref{fig:ABC_growth_frequency}) were also well reproduced compared to reference results.

Regarding divergence errors, VP maintains error levels below $20\%$ across all resistivities. For DI, however, divergence errors increase at low resistivity unless Dedner cleaning is applied (Figure \ref{fig:RF1_divB_err}). With cleaning, the errors remain within an acceptable range. Additionally, we assessed the stability of DI with divergence cleaning by introducing a significant initial monopole component in the magnetic field in the Roberts Flow I setup. In this scenario, Dedner cleaning effectively removed the errors, which did not significantly impact growth rates (Figure \ref{fig:Divergence_injection}).

To better track spatial divergence errors, we introduced additional monitoring quantities, which, in the low-resistivity regime, revealed regions not captured by conventional diagnostics (Figure \ref{fig:RF1_error_metrics}). We also established and tested the overwinding metric - a criterion for the onset of a numerically unresolved dynamo regime in Roberts Flow I (Eq. \ref{OW_definition}).

Furthermore, we performed several adiabatic cosmological runs (without subgrid modeling) with zero resistivity. These runs produced reasonable magnetic field, density, and velocity distributions (Figure \ref{fig:Cosmo_DI_IMHD_B_rho_v}) but exhibited large divergence errors, particularly near density gradients (Figure \ref{fig:Cosmo_DI_IMHD_errors}). By introducing a constant resistivity term based on the overwinding metric, we significantly reduced divergence errors while largely preserving the magnetic field structure (Figure \ref{fig:Cosmo_DI_OW}). However, a notable drawback of using constant resistivity is the initial overdamping of the magnetic field magnitude (Figure \ref{fig:Cosmo_DI_OW_Brms_vs_z}). This issue could potentially be mitigated by an adaptive resistivity approach, which future work should explore to balance error correction with the preservation of physical dynamics.\\

In conclusion, both the DI and VP SPMHD implementations in \textsc{Swift} successfully reproduce the action of kinematic dynamo in controlled environments such as in the Roberts flow I and ABC flows. We demonstrated that poor numerical behaviour at high Reynolds numbers can be identified using the overwinding trigger and divergence error metrics. Additionally, the code qualitatively reproduces cosmological simulations, solving the MHD equations in an expanding universe without subgrid physics. Introducing a constant resistivity term based on the overwinding metric effectively reduced divergence errors over the course of the cosmological simulation.
To further improve the implementations, an adaptive artificial resistivity scheme, dynamically adjusted based on local overwinding, could be implemented. This approach would better control divergence errors and enhance the field's smoothness while minimizing the risk of excessive magnetic field damping. 
\section*{Data Availability}
The SWIFT simulation code is entirely public, including the examples presented in
this work. It can be found alongside an extensive documentation on the website
of the project: \url{www.swiftsim.com}.
The {\sc Pencil Code} \citep{2021JOSS....6.2807P}, is freely available on
\url{https://github.com/pencil-code}.
The simulation setups and corresponding input
and reduced output data for the {\sc Pencil Code} runs are freely available on
\url{http://norlx65.nordita.org/~brandenb/projects/Roberts-Flow-Test}.



\bibliographystyle{mnras}
\bibliography{main} 


\bsp	
\label{lastpage}
\end{document}